\documentclass[aps,prl,superscriptaddress,floatfix]{revtex4}
\usepackage{graphicx}
\usepackage{epsfig}
\usepackage{bm}
\usepackage{thumbpdf}
\usepackage[colorlinks=true,linkcolor=black]{hyperref} 
\usepackage{amsfonts}
\usepackage{times}     
\usepackage{float}
\usepackage{indentfirst}   %
\usepackage{amsmath}
\usepackage{epstopdf}
\usepackage{textcomp}
\usepackage{tocvsec2}
\usepackage{epsfig}
\usepackage{xcolor}

\begin{document}
\title{Message spreading in networks with stickiness and persistence: Large clustering does not always facilitate large-scale diffusion}

\author{Pengbi Cui}
\affiliation{Institute of Computational Physics and Complex Systems, Lanzhou University, Lanzhou Gansu 730000, China and Key Laboratory for Magnetism and Magnetic Materials of the Ministry of Education, Lanzhou University, Lanzhou Gansu 730000, China}

\author{Ming Tang}
\affiliation{Web Sciences Center, University of Electronic
Science and Technology of China, Chengdu 610054, China}

\author{Zhixi Wu}\email{wuzhx@lzu.edu.cn}
\affiliation{Institute of Computational Physics and Complex Systems, Lanzhou University, Lanzhou Gansu 730000, China and Key Laboratory for Magnetism and Magnetic Materials of the Ministry of Education, Lanzhou University, Lanzhou Gansu 730000, China}

\begin{abstract}
Recent empirical studies have confirmed the key roles of complex contagion mechanisms such as memory, social reinforcement, and decay effects in information diffusion and behavior spreading. Inspired by this fact, we here propose a new agent--based model to capture the whole picture of the joint action of the three mechanisms in information spreading, by quantifying the complex contagion mechanisms as stickiness and persistence, and carry out extensive simulations of the model on various networks. By numerical simulations as well as theoretical analysis, we find that the stickiness of the message determines the critical dynamics of message diffusion on tree-like networks, whereas the persistence plays a decisive role on dense regular lattices. In either network, the greater persistence can effectively make the message more invasive. Of particular interest is that our research results renew our previous knowledge that messages can spread broader in networks with large clustering, which turns out to be only true when they can inform a non-zero fraction of the population in the limit of large system size.
\end{abstract}

\maketitle

Over the last few years, many empirical works~\cite{contagion,decay3,science,attention,contagion2,origin1,origin2} or practical model~\cite{zhou,model2} have identified the strong relevance of complex contagion mechanisms such as memory effect, social reinforcement and decay effects to information diffusion or behavior spreading. On account of memory effect, the previous contact activities can affect the current spreading process~\cite{memory,attention}. Specifically, individual's selection of message items can be naturally expedited by the increasing frequencies of the same choices of other people if they find the items interesting or crucial enough~\cite{science,zhou,decay}. This is usually interpreted as the results of social reinforcement~\cite{rein1,model2,theory2}. On the other hand, there are an increasing amount of new messages an individual is facing every day in modern real life, whereas the attention and processing abilities of people are finite and saturated~\cite{attention,decay,attention1}. The novelty of a message usually trend to fade with time and hence the attention people pay to it, which is normally described as decay effects~\cite{burst2,attention,contagion,decay}. It is shown that the social reinforcement effect could be weakened or even counterbalanced by decay effects~\cite{attention,decay,contagion2}.

Although the competition between social reinforcement and decay effects has been emphasized and used as a guideline to measure the natural time scale that attention fades away~\cite{decay}, to our best knowledge few works have been attempted to model explicitly the competition and memory effect, and study deeply how it shapes the spreading of information on complex networks. Here we want to point out that the three mentioned effects in information spreading are quite different from those have been considered in the studies on Naming Game (NG) and Category Game (CG), since either NG or CG is a two-step multi-state negotiation process~\cite{ng1,ng2,cg1,cg2}, whereas information spreading is not. First, herein memory effect performs as the storing of the times of contact of people with recipients of information~\cite{science,zhou}, rather than the possible words (or names) for the object (or a category) in NG (or CG)~\cite{ng1,ng2,cg1,cg2}. Second, decay effect in information spreading reflects the decay of people's interest or attention in a message owing to the competition with other news or stories~\cite{attention,decay}, contrary to the NG (CG) in which it means the decrease of the number of different words used in the system (or average number of words per category)~\cite{ng1,ng2,cg1,cg2}. Third, unlike the phenomenon that an hearing would have more opportunities to add (or remain) one word only if more selected speakers try to transmit the same one to it~\cite{ng1,ng2,cg1,cg2}, the reinforcement effect in information diffusion indicates the more simple situation that the more neighbors adopting the message, the higher likelihood an individual following them~\cite{science,contagion2}.

Next, the big challenge we are confronted with is the possibility of modeling and studying the message spreading along with both social reinforcement and decay effects based on the memory effect. Recent researches~\cite{origin1} have shown that the variation in the ways that different information spread is attributed to not only the stickiness -- the probability of information adoption is mainly dominated by the first few exposures~\cite{origin2, origin1}, but also the persistence -- the relative extent to which more repeated exposures to the message continue to have durative effect. Similar results especially the exposure response behaviors were also confirmed by a lot of empirical studies~\cite{contagion2,origin2,attention}. The two mechanisms, stickiness and persistence, thus enable us to quantitatively study the joint action of the three effects together.

At the same time, the structures of complex social systems can be characterized by complex networks, on which many spreading activities may take place, ranging from the spreading of epidemics~\cite{tang1,tang2,tang3,tang4}, the diffusion of behaviors and news~\cite{science,zhou}, to the promotion of technique innovations~\cite{innovation}, etc. Consequently, motivated by the empirical studies~\cite{science,zhou,attention,contagion2,origin1,origin2,decay} mentioned above, we propose a new agent-based model offering an opportunity to explore the impact of social reinforcement and decaying effects quantified by stickiness and persistence on the message (information) diffusion on various networks. In the presence of strong decay effects, we find that a message is more likely to outbreak (i.e., it can reach a non-zero fraction of the population in the thermodynamic limit) on the tree-like networks such as scale-free (SF) networks and Erd{\H{o}}s-R{\'e}nyi (ER) random networks rather than on the regular lattices (RLs). Specifically, a message can spread broader in the RLs than that in the tree-like networks only if it can outbreak. The critical behaviors of the diffusion process can be reasonably estimated by the bond-percolation theory considering spatial correlations of the underlying networks through which message diffuses. In addition, we develop a verification approximation, whose solutions confirm well the non-negligible role of the dynamical correlations between transmission events in the RLs.

\section*{Results}
\label{results}
Here, we first carry out extensive simulations for the agent-based model of message diffusion on square lattice. We then compare the simulation results with the predictions from the analytical bond percolation theory and verification method involving time correlations of the spreading events. Finally we extend our model and analytical methods to other networks such as RLs, SF networks, and ER networks to validate the robustness of our findings.

\textbf{Message diffusion on square lattice.}
\label{square}
We first consider the message diffusion on a square lattice of size $N=L\times L$ with periodic boundary conditions. The message starts spreading from the center node (selected as the seed), while all the others are in the susceptible state (i.e., they hear nothing about the message).

\begin{figure}[H]
\centering
\includegraphics[width=\textwidth]{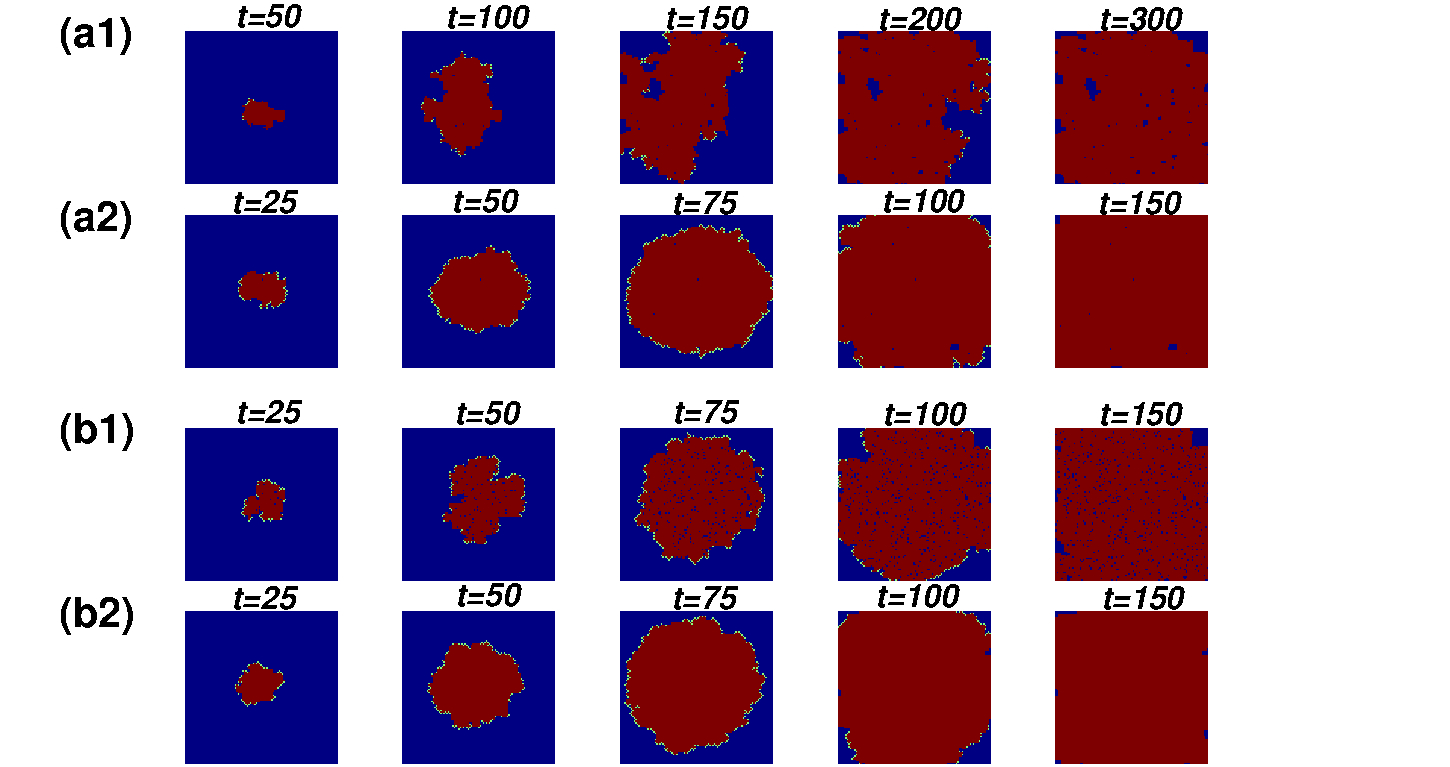}
\caption{\textbf{The time evolution of spatial patterns,} for two different values of stickiness (a1) $a=0.40$, (a2) $a=0.45$ (bottom panel) where $b=0$; and for two values of persistence (b1) $b=-1.00$, (b2) $b=1.00$ (bottom panel) where $a=0.45$. Red sites represent recovered or alerted nodes, bright green sites represent infected ones, and blue sites denote susceptible nodes. Other parameters are chosen as $n_{s}=2$ and $N=101\times 101$.}
\label{spatial}
\end{figure}

To intuitively grasp the roles of stickiness and persistence, we begin by presenting the time evolution of spatial patterns of message spreading in Fig.~\ref{spatial}. The message with stickiness $a=0.40$ (see the Methods for the precise definitions of $a$, $b$ and other parameters) spreads in an irregular manner~(see Fig.~\ref{spatial}(a1)). By comparison, the message with a slightly stronger stickiness $a=0.45$ diffuses outward to susceptible areas in a quasi-circular manner with a broader rim of informed (infected) individuals (see Fig.~\ref{spatial}(a2)). This indicates that messages with different strength of stickiness could give rise quite different spreading patterns and behaviors. Figs.~\ref{spatial}(b1)~(b2) and Supplementary Fig.~S1 show that the persistence $b$ also affects considerably the whole spreading size, by governing the number of the isolated susceptible islands (blue domains surrounded by red areas, the emergence of these islands arises from the fact that continually increasing number of infected neighbors fail to infect those individuals owing to small persistence). The above arguments suggest that the stickiness and persistence have great but different influences on the spreading of message.

\begin{figure}[H]
\centering
\includegraphics[width=\textwidth]{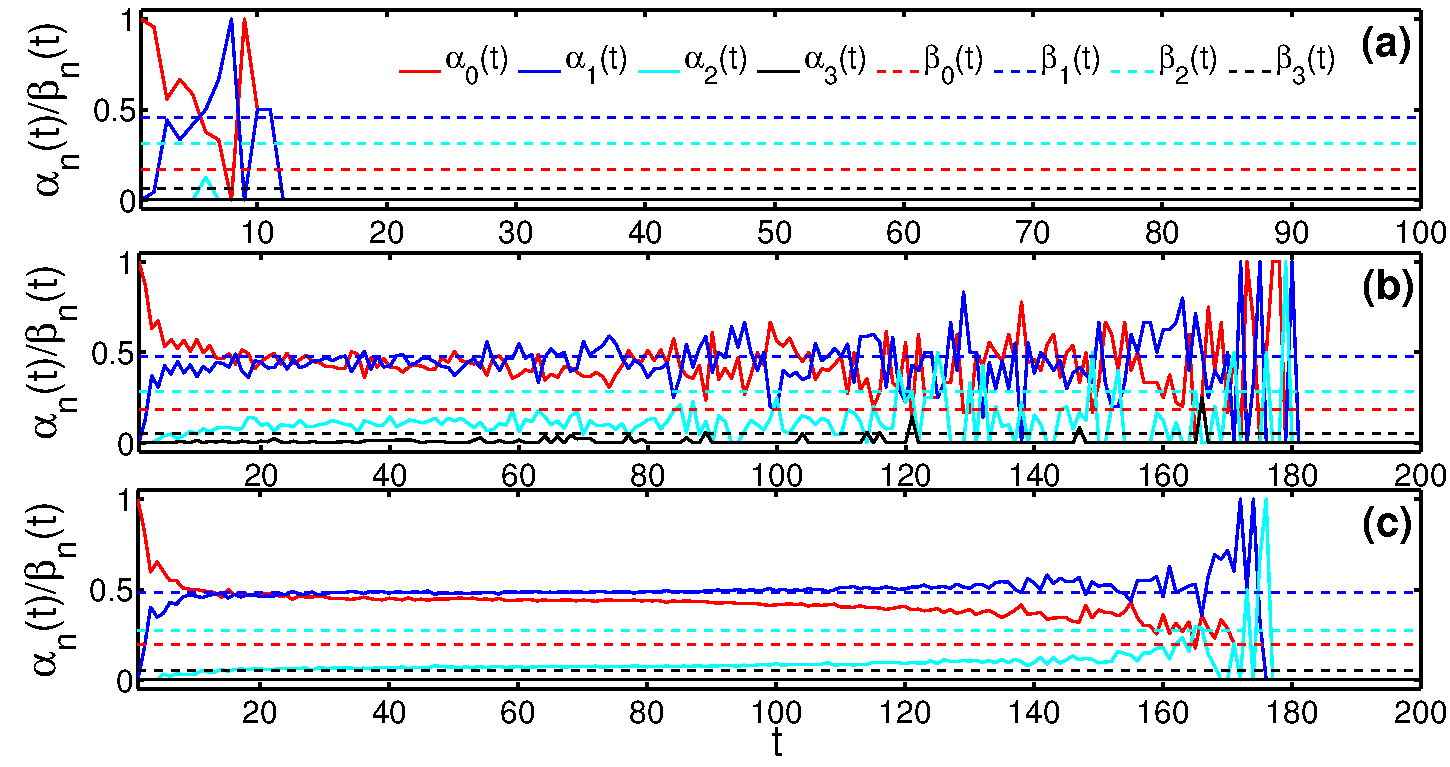}
\caption{\textbf{The evolution of proportions of the transmission events.} The parameters are chosen at (a) subcritical point $a=0.20$, $b=0.20$; (b) critical point $a=0.35$, $b=0.20$; and (c) supercritical point $a=0.43$, $b=0.20$.}
\label{alround}
\end{figure}

We next explore the critical behaviors of message diffusion for various stickiness and persistence, by providing a quantitative assessment of the message burstiness. By means of the theory of non-equilibrium phase transition in statistical physics which has been successfully generalized to study the epidemic dynamics, previous studies~\cite{variance1,variance2} on spreading dynamics have shown that the fluctuation of the order parameter is divergent at the critical point. We thus use the procedure proposed in~\cite{variance2} to numerically determine the critical areas. To be more specific, a series of variabilities $v(a,b)$ are firstly obtained as
\begin{equation}\label{eq:variance}
v(a,b)=\frac{\sqrt{<\rho^{2}_{R}(a,b)>-<\rho_{R}(a,b)>^{2}}}{<\rho_{R}(a,b)>},
\end{equation}
where $\rho_{R}$ and $v(a,b)$ denote, respectively, the density of recovered individuals in the population and the relative standard deviation (RSD) at parameter point $(a,b)$. There exists a maximum variability $v_{max}(b)$ for each value of $b$ when varying $a$ from $0$ to $0.5$, and the values of  $v_{max}(b)$ with $b\in[-1~1]$ can be used as the numerical estimation of the threshold position.

Although a bond-percolation process can be mapped to the SIR model~\cite{prl,bond}, its extension to the current model is not straightforward. First, in our model, the transmission probability that a susceptible individual approves the message varies with the times he has received it from his infected neighbors (i.e., the number of informed neighbor he has had). Second, the time correlations between different transmission events $E_{n}$~\cite{prl} (the transmission event $E_{n}(t)$ represents that an individual, who has received the message at least once before, successfully approves the message when he has received it another $n$ times until time $t$). To confirm the existence of the time correlation, in Fig.~\ref{alround} we compare the dynamic proportions of the four transmission events ($\alpha_{n}(a,b,t)=\frac{\omega_{n}(a,b,t)}{\sum^{\langle k\rangle -1}_{m=0}\omega_{m}(a,b,t)}$) obtained from numerical simulations with those predicted by the bond percolation considering spatial correlations ($\beta_{n}(a,b,t)=\frac{q_{n}T_{n}(a,b)}{\sum^{\langle k\rangle -1}_{m=0}q_{m}T_{m}(a,b)}$ and see the Methods section for definitions of $q_{n}$ and $T_{n}(a,b)$) for three parameter points $(a,b)$. Here, $\omega_{n}(a,b,t)$ represents the occurrence frequency of $E_{n}(t)$ obtained from numerical simulations, and $\langle k\rangle$ denotes the average degree of the networks. In Figs.~\ref{alround} (b) and (c), we see that $\alpha_{0}(t)$ ($\alpha_{2}(t)$) is greater (less) than $\beta_{0}(t)$ ($\beta_{2}(t)$) during the spreading process, whereas $\alpha_{1}(t)$ ($\alpha_{3}(t)$) is equal or close to $\beta_{1}(t)$ ($\beta_{3}(t)$). The reason is that the existence of time correlations of the transmission events $E_{n}(t)$ can determine whether the subsequent events $E_{m}(t)$ ($m>n$) happen or not in the spreading process. If $E_{n}(t)$ does not happen, the events $E_{m}(t)$ ($m>n$) will probably happen; otherwise $E_{m}(t)$ ($m>n$) will never happen since an informed individual transmits the message only one time and then becomes recovered (i.e., completely ignores the message) forever. Consequently, $E_{0}$ ($E_{2}$) contributes more (less) than predicted by percolation theory to the spreading course (also see Supplementary Fig.~S2). To overcome the challenge, we develop a verification approximation involving the time correlations of the transmission events, besides the spatial correlations originating from the spatial structure of the lattice~\cite{prl}. It is necessary to mention that the discrete-time synchronous transmission of the message enables us to avoid concerning about the additional synergistic effect~\cite{synergy}. Moreover, Figs.~\ref{alround}(b) and (c) show that $\alpha_{n}(a,b,t)$ is dynamically stable, which shows that this correlation always exists. It allows us to adopt average values of the four indices $\alpha_{n}(a,b,t)$ at the critical regions for the verification approximation.

\begin{figure}[H]
\centering
\includegraphics[width=\textwidth]{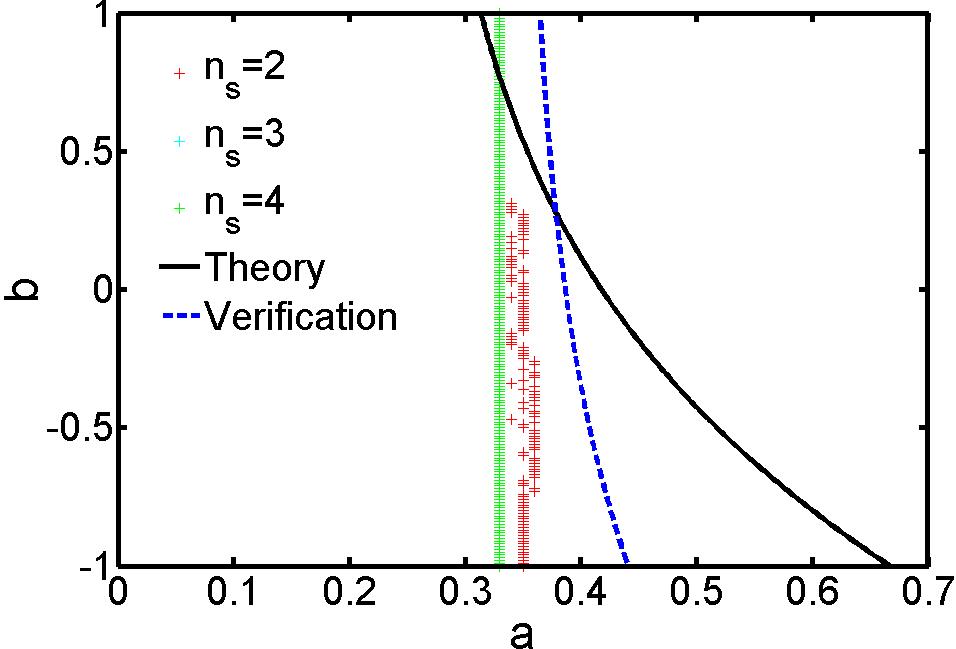}
\caption{\textbf{The phase diagram of the message spreading.} The real numerical critical boundaries (crosses) are obtained by Eq.~\eqref{eq:variance} for various $n_{s}$ on square lattice of size $N = 101\times 101$. For comparison, analytical boundary (black line) and verification boundary (dash black line)  for $n_{s}=2$ are also shown (the cases for $n_{s}=3,~4,~5$ do not allow us to analytically identify the critical lines). Herein, we select a narrow parameter ranges $b\in[-1,~1]$ and $a\in[0.32,~0.36]$ containing the numerical critical boundary for the calculation of verification threshold (see more detailed method in the Methods section).}
\label{latthre}
\end{figure}

Based on the proposed methods in the Methods section and Eq.~\eqref{eq:variance}, we yield both the analytical prediction and the verification threshold for $n_{s}=2$, plus the numerical results for various $n_{s}$, as depicted in Fig.~\ref{latthre}. We note that the numerical thresholds stay at $a\approx 0.32$, which is mainly determined by the stickiness of message (i.e., the parameter $a$), regardless of the values of $b$ and $n_{s}$. This means that most informed individuals are actually infected by their first one or two infected neighbors (also see Supplementary Figs.~S2--S6). Furthermore, the numerical estimations are fairly reproduced by the bond-percolation theory. Comparing the analytical boundary, the verification approximation involving both spatial and time correlations gives a higher accurate estimation than the bond-percolation method considering only spatial correlations (the dashed black line is clearly closer to the numeric markers than the black solid line is). 

\textbf{Message diffusion on regular lattice networks and regular random networks.}
\label{hm}
Centola's work~\cite{science} concludes that social behaviors can spread farther and faster across clustered-lattice networks than across corresponding regular random networks (RRNs), owing to the strong social reinforcement induced by clustered ties. RRNs are networks that all nodes have exactly the same degree while links are randomly distributed among nodes, avoiding self-connections and multiple connections. To check  whether the findings by Centola are still fulfilled for information diffusion, we further investigate our model defined on the RLs (Hexagonal network and Moore network) and the RRNs.

\begin{figure}[H]
\centering
\includegraphics[width=\textwidth]{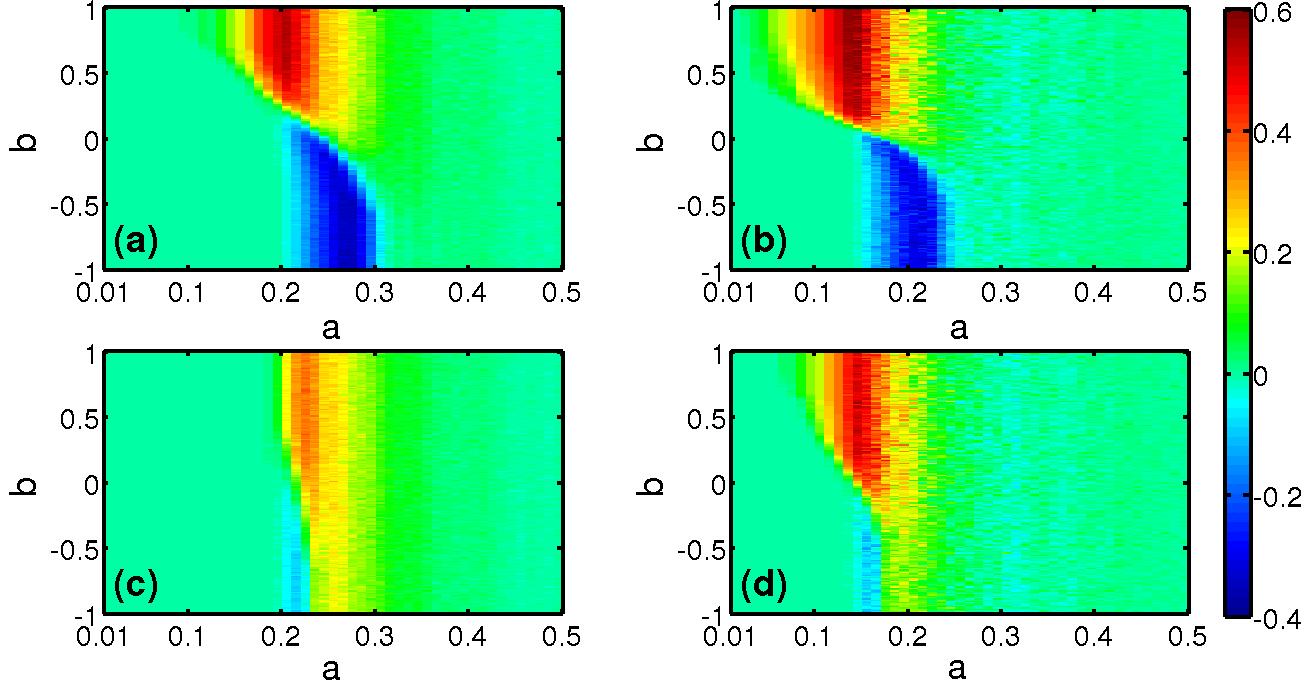}
\caption{\textcolor{red}{\textbf{The differences of spreading sizes between RLs and RRNs networks ($\rho_{RL}-\rho_{RRN}$).} (a,b) $n_{s}=2$, (c,d) $n_{s}=3$. The average degrees of the networks are $\left\langle k\right\rangle=6$ (a,c), and $\left\langle k\right\rangle=8$ (b,d), respectively. $\rho_{RL}$ ($\rho_{RRN}$) represents the recovered density of population (i.e., spreading size) on RLs (RRNs).}}
\label{lhgap}
\end{figure}

To get a comparison, we present in Fig.~\ref{lhgap} the differences of the final size of recovered population on the two networks with the same average degree. The blue areas characterize the parameter regions where the conclusion of Centola's experiment (that the information spreads farther across the RLs than across the corresponding RRNs) does not hold. The violation is attributed to the presence of strong decay effects ($b<0$) in the vicinity of the critical regions. Specifically, the outbreaks of message can happen more easily in the RRNs than that in the RLs for negative persistence owing to that strong decay effects outcompetes the weak reinforcement effect (also see Supplementary Fig.~S7, strong reinforcement effect (decay effects) is reflected by large stickiness and/or positive persistence (negative persistence) in our model). As $a$ and $b$ get larger, things turn out differently, the message is able to seize a larger population in the RLs, which is accordant with the anticipation of Centola's experiment. This means that high level of clustering created by redundant ties that linked each node's neighbors to one another in the RLs strengthens the reinforcement effect, and hence facilitates the diffusion of the message~\cite{science}. Moreover, larger $n_{s}$ improves the performance of stickiness in facilitating the message diffusion, making social reinforcement effects be the most prominent for the RLs. Consequently, the blue areas shrink with increasing $n_{s}$. Thus, the above differences investigated indicate that network topology and $n_{s}$ (which can be regarded as one of the intrinsic characteristics of the message) simultaneously determine the effects of stickiness and persistence on the spreading dynamics.

\begin{figure}
\centering
\includegraphics[width=\textwidth]{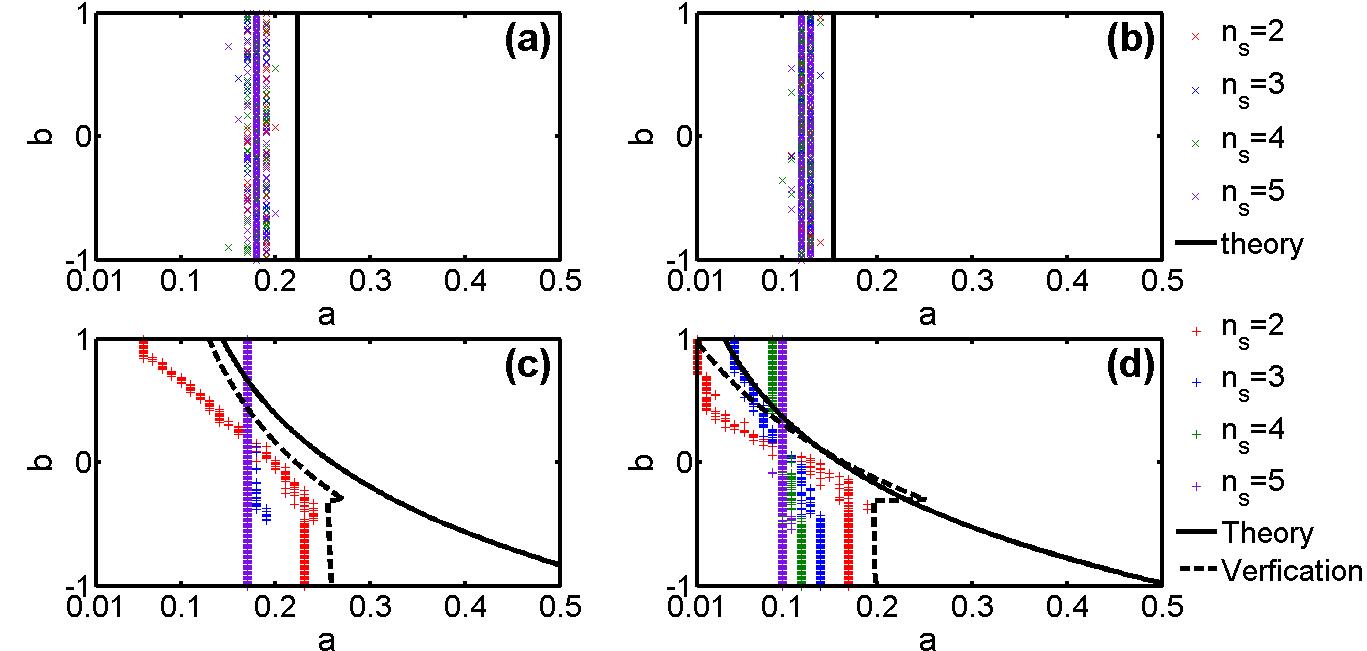}
\caption{\textbf{The phase diagrams of the message spreading on RLs and RRNs.} The underlying networks are (a) RRN with $\left\langle k\right\rangle=6$, (b) RRN with $\left\langle k\right\rangle=8$, (c) Hexagonal lattice, and (d) Moore lattice. The predictions from the bond-percolation theory (solid lines) and verification approximations (dashed lines) for $n_{s}=2$ are illustrated to compare with the simulation data (markers) for various $n_{s}$. To get the verification trajectory on Hexagonal lattice (and Moore lattice), we select two parameter regions containing the numerical thresholds (see more detailed method in the Methods section). One region is $b\in[-0.3~1.0]$ and $a\in[0.06~0.24]$; the other one is $b\in[-0.3~1.0]$ and $a\in[0.01~0.19]$ in Hexagonal lattice. One region is $b\in[-1.0~-0.3]$ and $a\in[0.23~0.24]$; the other one is $b\in[-1.0~0.3]$ and $a\in[0.15~0.18]$ in  Moore lattice.}
\label{rhthre}
\end{figure}

Using Eq.~\eqref{eq:variance}, we obtain the numerical thresholds for various values of $n_{s}$. According to the methods described in the Methods section, we can yield the analytical (theoretical) thresholds of message diffusion on both the RRNs and the RLs, in addition to the verification thresholds on the RLs for $n_{s}=2$ (Supplementary Fig.~S10 and Fig.~11 show that the time correlations between the transmission events in Hexagonal lattice and Moore lattice are noticeable). In the case of RRNs (Figs.~\ref{rhthre}(a) and (b)), we observe that the positions of the critical boundaries are mainly determined by both the stickiness and $\left\langle k\right\rangle$ instead of the persistence and $n_{s}$, on account of weaker social reinforcement~\cite{science,prl} resulting from the low clustering coefficient. Unlike the case of RRNs, the theoretical analysis by means of bond-percolation theory gives rather accurate predications of the position of the threshold with strong persistence, but not for negative persistence (i.e, with the presence of strong decay effects) on the RLs. When the underlying networks for message diffusion are Hexagonal lattice and Moore lattice, more available edges for message spreading will further strengthen the role of persistence in message diffusion, especially with stronger social reinforcement (positive $b$). That enables all transmission events involve in the diffusion (see Supplementary Figs.~S10--S13), so that $\beta_{n}(a,b,t)$ gets close to $\alpha_{n}(a,b,t)$ (see Supplementary Figs.~S10 and S11). On the other hand, the trajectories obtained by verification approximation are in good agreement with the simulations, from which one can conclude that the effect of time correlations of transmission events is indeed general on the RLs. Additionally, the message steps forward to arrive in half of the neighbors of the same host on the RLs by flowing through almost $\frac{\left\langle k\right\rangle}{2}$ edges connecting it. This makes the message with positive persistence ($b>0$) outbreak more easily in the presence of denser local connections, and the persistence thus imposes a greater influence on outbreaks of message on RLs with larger average degree for small $n_{s}$ ( $\frac{n_{s}}{\left\langle k\right\rangle}<\frac{1}{2}$). The message also has reached a saturation state when the subsequent events $E_{i}$ ($i>\frac{\left\langle k\right\rangle}{2}$) happen (see Supplementary Figs.~S14--S21). Also in RLs, higher $n_{s}$ limits the effect of persistence, and the phase transitions are determined by not only the topologies of the networks but also the stickiness and persistence of the message.

In addition, the reinforcement effect begins to work as the message is bursting and prevailing on both the RLs and RRNs. Therefore, the results for positive persistence near the thresholds (see Supplementary Figs.~S12--S24) elucidate the actual phenomenon ``Three men make a tiger" (or ``A lie, if repeated often enough, will be accepted as truth"), where the majority do not believe the message as a truth until at least three neighbors have tried to transmit this message to them.

\textbf{Message diffusion on SF networks and ER networks.}
\label{eb}
To further check the robustness of our above findings, we finally investigate our message diffusion model on SF and ER networks with $\langle k\rangle=6, 8, 10, 12$. Since two or more transmission events fail to last over the long time at the critical points~(see Supplementary Fig.~S24), we do not take the time correlations into consideration in theoretical analysis. Compared with the simulation data, the analytical results for SF networks and ER networks with different average degrees show that the theoretical approaches are already sufficient to give fairly precise expressions of the outbreaks of message completely determined by the stickiness $a$~(see Supplementary Fig.~S25). Nevertheless, the persistence also partly boosts its impact on the size of message diffusion as $\left\langle k\right\rangle$ gets larger (see Supplementary Fig.~S26).

\begin{figure}[H]
\centering
\includegraphics[width=\textwidth]{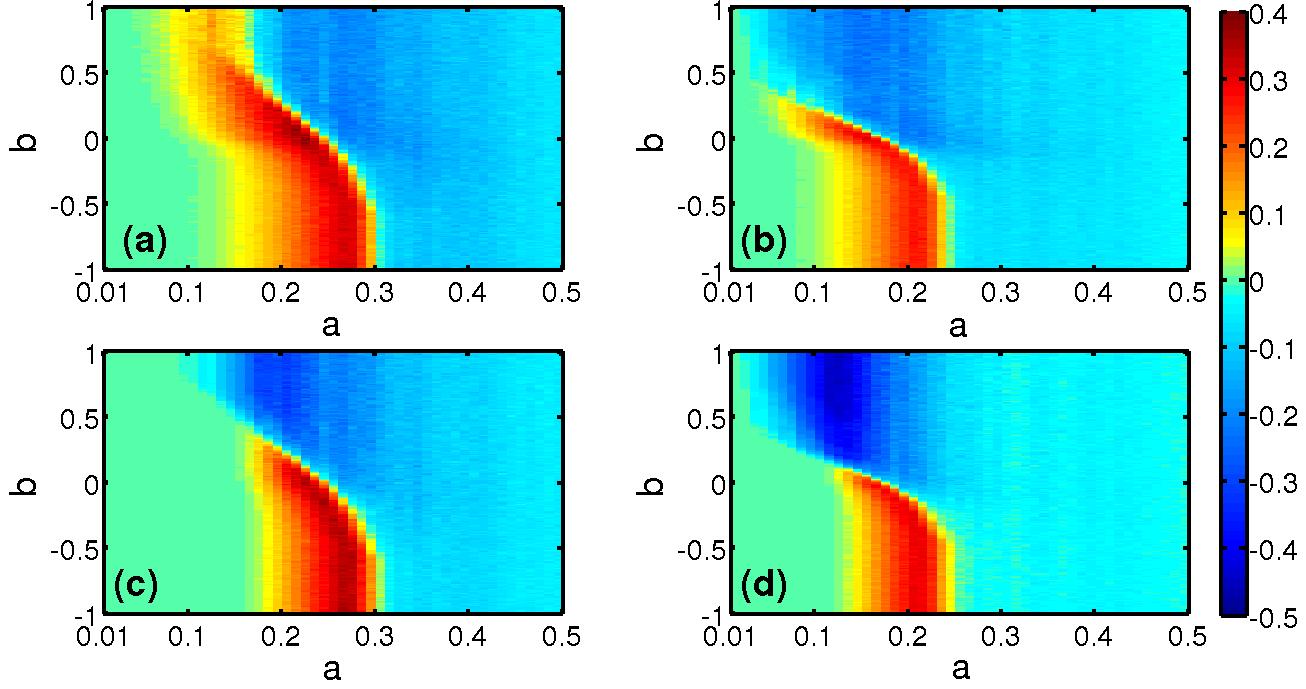}
\caption{\textbf{The differences of spreading sizes between the RLs and SF (ER) networks ($\rho_{SF/ER}-\rho_{RL}$).} (a,b) the difference $\rho_{SF}-\rho_{RL}$, (c,d) the difference $\rho_{ER}-\rho_{RL}$. (a,c) The average degrees of the networks are $\left\langle k\right\rangle=6$ (a,c),  and $\left\langle k\right\rangle=8$ (b,d), respectively. Here, $\rho_{SF}$ ($\rho_{ER}$, $\rho_{RL}$) represents the recovered density in SF networks (ER networks, RLs). Other parameter: $n_{s}=2$.}
\label{brhmgap}
\end{figure}

Fig.~\ref{brhmgap} displays evident differences of the final spreading sizes on the RLs and SF (ER) networks with the same degree. The spreading sizes are larger in SF (ER) than those in the RLs at the parameter regions where the message has already outbroken on SF (ER) but not yet on the RLs, owing to the hubs and shorter shortest paths in SF (ER)~\cite{shortest} (see Supplementary Fig.~S25 for the critical boundaries of message on SF and ER). Just for strong persistence, denser local connectivity of the RLs can make an invasion take place more easily. As $a$ and $b$ increase beyond the critical boundaries displayed in Fig.~\ref{rhthre}, the message can capture a larger population again on the RLs, despite of the existence of hubs the short characteristic path length in the SF (ER) networks. The reason is that very smaller clustering coefficient gives rise to weak social reinforcement effect~\cite{science,report}, which again leads to weak performance of persistence in promoting the spread of message on the SF (ER) networks. Our results indicate that the role of reinforcement effect is more important than that of hubs or shortest paths in facilitating message spreading only when the message outbreaks on the RLs. Otherwise, especially for the presence of decay effects (negative persistence) the facilitation of hubs and shortest paths to the diffusion on SF (ER) cannot be neglected. In addition, the results for positive persistence on SF and ER networks can also be treated as evidence of the mechanism "Three men make a tiger" (see Supplementary Fig.~S27 and Fig.~S28).

\section*{Discussion}
\label{con}
In conclusion, taking into account social reinforcement and decay effects based on memory effect in reality, we have proposed a new agent-based message spreading model with stickiness and persistence, and carried out extensive computer simulations of our model on various types of networks. By means of the relative standard deviation method and the bond percolation theory involving the spatial correlations, we are able to determine numerically and analytically the positions of critical boundaries. Moreover, the remarkable accuracy of verification approximation involving the time correlations between different transmission events validates the wide existence of such correlations for the message diffusion on regular lattices.

Our preliminary results show that in RLs, the persistence depends greatly on the position of inflection point $n_{s}$ and average degree $\left\langle k\right\rangle$ of the underlying networks, and begins to play a pivotal role in the spreading process with increasing $\left\langle k\right\rangle$ owning to the emergent large clustering coefficient~\cite{science}. Stronger social reinforcement arising from larger clustering coefficient leads readily to stronger infectivity of the message, which can invade a great number of susceptible individuals in the RLs, confirming the conclusion of Centola's work. By comparison, in tree-like networks such as RRNs, SF and ER, the critical thresholds of message diffusion are only dominated by the stickiness, and both the hubs and the short characteristic path length facilitate the outbreaks of message in the presence of decay effects. It worth emphasizing that the results presented in this paper has successfully substantiated the phenomenon ``Three men make a tiger" (or "A lie, if repeated often enough, will be accepted as truth").

Placing our study in the context of social media, the hub nodes actually play a role of broadcasters, advertisements and so on, which are very important for the large scale spreading of information. However, the intrinsic contents of messages and their adaptation to hosts are extremely relevant in determining the message diffusion. In particular, the results for the message diffusion on the SF and ER networks constitute a proof that exposure to mass media can favour the outbreaks of behaviors, news or messages, despite the decay effects, only if the stickiness of the message is large enough. On the other hand, the RLs are efficient in taking advantage of social reinforcement effects to promote the global spreading of the message, owning to more cluster ties (or local pressures) that function as form of 'initial groups' or 'small groups' through interpersonal communication or 'machine-interpersonal communication'~\cite{fed}. It is also verified by our research that the local, personal communication is irreplaceable to lead to propagation of message despite of the developed media industry today, from the perspective of communication.

Recent studies have started considering the memory effect~\cite{science,zhou}, social reinforcement and decay effects~\cite{attention} in information spreading. The mechanisms were yet investigated in isolation. Our work is the first attempt to account explicitly for the three key mechanisms together, and to evaluate the joint action of them by quantifying their effects as stickiness and persistence. It provides a quantitative guideline for future social experiments for message spreading.

In reality, the ways in which information spread may be very complicated. In the present study, we do not capture the difference of individual-level preference~\cite{leader} that might have also influenced their decisions to adopt one message. For example, in an online social network such as Twitter, individuals may prefer different hashtags, and significant variations in the ways that the hashtags on different topics spread were observed~\cite{origin2}. In addition, one need to concern about the diverse cultural and societal backgrounds~\cite{culture} which would lead to the different styles by which individuals contact with the medium, or even hamper communications among different groups of members~\cite{fed}. Moreover, the volatilities of complex contagion of controversial topics, psychological status of individuals~\cite{media} reveal that the status of the individuals in the communication systems are time dependent, which should be addressed in future research.

\section*{Methods}
\textbf{Message spreading model with stickiness and persistence.}
The message spreading model is implemented on a network consisting of $N$ nodes and $E$ edges, where the nodes represent the individuals in a population and the edges the social interactions among them, through which information propagates. Each individual is allowed to be in one of three states at each time step: (i) Susceptible (or uninformed) state---the individual has not yet heard the message or is aware of the news but not willing to transmit it. (ii) Infected state---the individual catches the message and forwards it to all his nearest neighbors. (iii) Recovered state---the individual will never transmit the message any more after having transmitted it once before.

\begin{figure}[H]
\centering
\includegraphics[width=\textwidth]{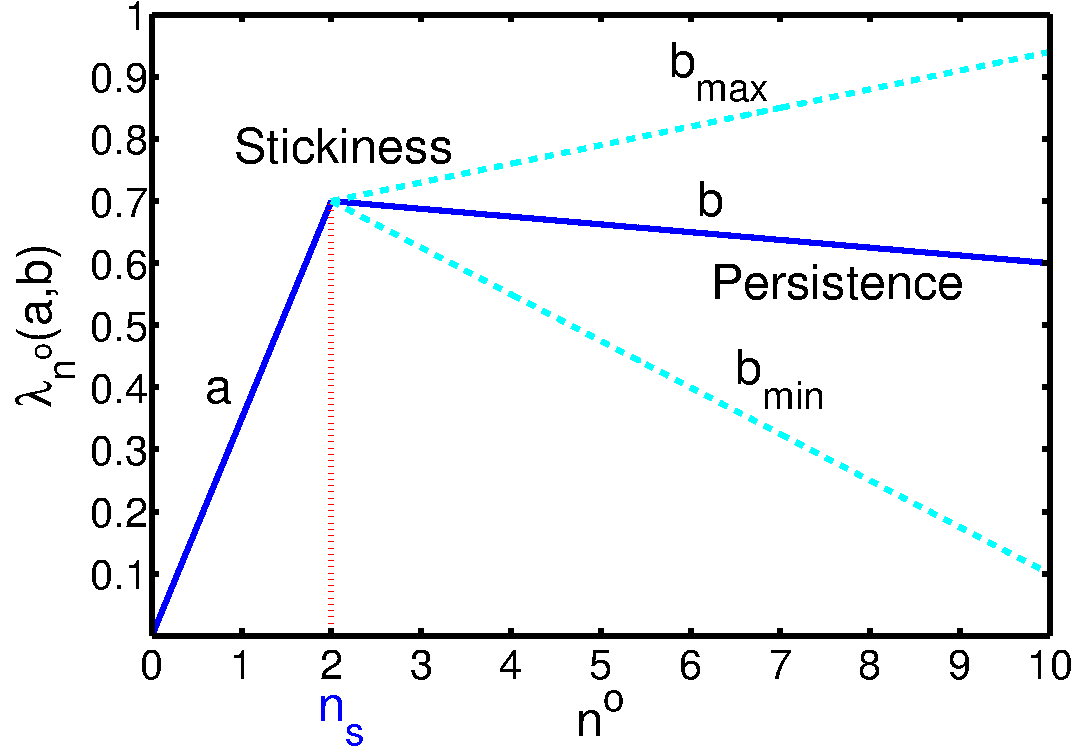}
\caption{\textbf{The adopting probability of message as a function of $n^{o}$.} The degree of stickiness and persistence are  quantified as $an_{s}$ and $b$, respectively. $n_{s}$ denotes the position of inflection point.}
 \label{model}
\end{figure}

Specifically, the information propagation is modeled in a probabilistic framework at individual level~\cite{zhou,model2}. A susceptible individual $i$ will adopt the message with a probability $\lambda_{n^{o}}(a,b)$, given that he has heard it from his informed neighbors $n^{o}$ times (i.e., $n^{o}$ informed neighbors he has had), plus the first time. In detail, $n^{o}=n+1$ when individual $i$ has owned at least one informed neighbors, otherwise $n^{o}=0$. Here $\lambda_{n^{o}}(a,b)$ is a linear piecewise function of the times he has received message from his informed neighbors:
\begin{align}
\label{eq:lambda}
	&\lambda_{n^{o}}(a,b) =
	\begin{cases}
	\min\{an^{o}, 1\}, \quad 0\leq n\leq n_{s};\\
	\min\{bn^{o}+n_{s}(a-b),1\}, \quad n_{s}< n\leq k_{i};
	\end{cases}\\ \notag
	&\text{and} \quad  \lambda_{n^{o}}(a,b)  =  0, \qquad \text{if} \quad \lambda_{n^{o}}(a,b)<0;	
\end{align}
 where $k_{i}$ is the degree of node $i$. $n_s$ is the inflection point beyond which persistence is the dominant factor for the infection probability of message, and the parameters $a$ (stickiness) and $b$ (persistence) characterize how $\lambda_{n^{o}}(a,b)$ change with $n^{o}$, as illustrated in Fig.~\ref{model}. Since empirical data~\cite{science,origin2,origin1} have shown that social reinforcement sets in such that initial exposures generally increase infection probability, the parameter $a$ should be non-negative when $n^{o}\leq n_{s}$. For $n^{o}> n_{s}$, the competition between social reinforcement and decay effects, characterized by the parameter $b$, will be taken into account for the message adoption. If the reinforcement is strong enough the individual will be more likely to adopt the message with increasing $n^{o}$ ($b>0$). Otherwise, even if many infected neighbors try to transmit the information to the focal individual $i$, the multiple exposures will lead to a decreased probability for the information adoption ($b<0$). In the present study, we set $b\in[-1,1]$ and $a\in(0, 0.5]$ so that the spreading dynamics of the message can be comprehensively investigated. For simplicity, we do not consider the diversity of individuals' response to the message, and all individuals behave identically with the same values of parameters $a$ and $b$.

We perform Monte Carlo (MC) simulations with synchronous updating of the states of all the individuals. Each MC step consists of the following three procedures: (i) All susceptible individuals decide whether or not to adopt the message with probability $\lambda_{n^{o}}(a,b)$; (ii) If an individual adopts the message, he will try to transmit what he has approved to all his nearest susceptible neighbors in the next step, and then becomes recovered immediately; (iii) Otherwise, the susceptible individuals will wait to repeat the procedure (i) in the following MC steps. The above elementary spreading processes are repeated $T^{'}_{S}=500$ steps until there are no infected individuals anymore in the population.

\textbf{Theoretical analysis of the model.}
For the occurrence probability of transmission event $E_{n}$, $T_{n}(a,b)= 1-e^{-\lambda_{n+1}(a,b)\tau}\notag = 1-e^{-\lambda_{n+1}(a,b)}$ with $\tau =1$~\cite{newman}. We consider the spatial correlations that affect the process of diffusion, but do not yet influence the critical behaviour of the message spreading~\cite{prl}. The transition point from susceptible to infected phase is determined by~\cite{prl, newman}
\begin{eqnarray}
T_{C}=\left\langle T\right\rangle,
\label{eq:threshold}
\end{eqnarray}
where $\left\langle T\right\rangle$ and $T_{C}$ are, respectively, the mean transmissibility and the critical topology-dependent bond-percolation threshold. On the other hand, the mean transmissibility can be gotten as
\begin{eqnarray}
\left\langle T\right\rangle & = & \sum^{\left\langle k\right\rangle -1}_{n=0}q_{n}T_{n}(a,b),
\label{eq:meanthreshold}
\end{eqnarray}
where $q_{n}=\binom{\left\langle k\right\rangle-1}{n}p^{n}(1-p)^{\left\langle k\right\rangle-n-1}$ is the probability that the recipient has other $n$ ($n=0, 1, 2, 3$) infected neighbors except for the one chosen beforehand when considering the spatial correlations ($p$ is the probability that one nearest neighbor of the focal individual is in infected state).

Combine Eqs.~\eqref{eq:lambda}, \eqref{eq:threshold}, \eqref{eq:meanthreshold}, and $T_{n}(a,b) = 1-e^{-\lambda_{n+1}(a,b)}$, the mean transmissibility for the discrete case reads as
\begin{align}
\label{eq:percolation2}
\left\langle T\right\rangle & =  \sum^{\left\langle k\right\rangle-1}_{n=0}q_{n}(1-e^{-\lambda_{n+1}(a,b)}) \notag \\
	  & =  \sum^{\left\langle k\right\rangle -1}_{n=0}q_{n}-\sum^{n_{s}-2}_{n=0}q_{n}e^{-\lambda_{n+1}(a,b)}-\sum^{\left\langle k\right\rangle -1}_{n=n_{s}-1}q_{n}e^{-\lambda_{n+1}(a,b)}   \\
	  & =  1-\sum^{n_{s}-2}_{n=0}q_{n}e^{-(n+1)a}-e^{-n_{s}a}\sum^{\left\langle k\right\rangle -1}_{n=n_{s}-1}\binom{\left\langle k\right\rangle -1}{n}e^{-(n-1)b}.\notag
\end{align}
Then we have
\begin{eqnarray}
\theta_{2}e^{-2a}+\theta_{1}e^{-a}+\theta_{0}=0,
\label{eq:lattper}
\end{eqnarray} where $\theta_{2}=\sum^{\left\langle k\right\rangle-1}_{n=1} \frac{(\left\langle k\right\rangle-1)!}{n!(\left\langle k\right\rangle-n-1)!} p^{n}(1-p)^{\left\langle k\right\rangle-n-1} e^{(1-n)b}$, $\theta_{1}=(1-p)^{\left\langle k\right\rangle-1}$ and $\theta_{0}=T_{C}-1=-\frac{1}{2}$. The positive root is selected as the theoretical prediction
\begin{equation}
a(b)=\ln(\frac{2\theta_{2}}{\sqrt{\theta^{2}_{1}-4\theta_{2}\theta_{0}}-\theta_{1}}).
\label{eq:lattre}
\end{equation}
In SF networks, the critical point $T_{C}$ beyond which the message can reach a finite faction of the population can be obtained~\cite{newman} as $T_{c}=\frac{\left\langle k\right\rangle}{\left\langle k^{2}\right\rangle -\left\langle k\right\rangle}$. Since RRNs and ER networks are generated by connecting randomly selected pair of nodes, $T_{c}=\frac{1}{\left\langle k\right\rangle-1}$. Due to the randomness of connections in these networks, each edge of a neighbor of one host can probably connected to any other $N-2$ individuals. Therefore, the transmission of message from an informed neighbor to the susceptible host (i.e., message flows through the edge in the system) will happen only if there is at least one infected individual in the left $N-2$ ones to transmit the message to the recipient. In other words, the probability that the message flow can reach the susceptible host through the edges connecting to the recipient is $p=\frac{1}{N-2}$ for the three networks (i. e., the RRNs, SF networks, and ER networks). Also, two or more transmission events fail to last over the long time at the critical points (see Supplementary Fig.~S8, Fig.~S9, and Figs.~S25), so the time correlations are ruled out in the theoretical analysis for the cases of RRNs, SF networks, and ER networks. In the regular lattices, $p=\frac{1}{\left\langle k\right\rangle-1}$ since each one has exact $\langle k\rangle$ specific neighbors. Specifically, $p=\frac{1}{3}\approx0.333$ for square lattice with von Neumann neighborhood, $p=\frac{1}{5}$ for Hexagonal lattice, and $p=\frac{1}{7}\approx0.143$ for the lattice with Moore neighborhood. The bond percolation threshold $T_{C}=\frac{1}{2}$ for a square lattice, $T_{C}=0.347$ for Hexagonal lattice, and $T_{C}=0.232$ for square lattice with Moore neighborhood~\cite{prl,tc}.

\textbf{Verification approximation of information threshold.}
In analogous to $q_{n}(a,b)$, we first set $Q^{'}_{n}(a,b) (n=0,~1,~2,~3)$. According to Eq.~\eqref{eq:meanthreshold}, $Q^{'}_{n}(a,b) (n=0,~1,~2,~3)$ can be calculated numerically by counting the relative number of successful attacks (i.e, transmission events) $\alpha_{n}(a,b)$ from infected neighbors to hosts for given values of $a$ and $b$ and equating this to $\alpha_{n}(a,b)=\frac{\sum_{t=0}^{T^{'}_{S}}\omega_{n}(a,b,t)}{\sum^{\langle k\rangle -1}_{m=0}\sum^{T^{'}_{S}}_{t=0}\omega_{m}(a,b,t)}=\frac{Q^{'}_{n}(a,b)T_{i}(a,b)}{<T(a,b)>}$. In other words, $\alpha_{n}(a,b)$ represents numerically the proportion of $E_{n}$ occurring at parameter point $(a,b)$. Furthermore, $Q^{'}_{n}(a,b)$ can be normalized as
\begin{equation}
Q_{n}(a,b)= \frac{Q^{'}_{n}(a,b)}{\sum^{\langle k\rangle -1}_{m=0}Q^{'}_{m}(a,b)}= \frac{\frac{\alpha_{n}(a,b)<T(a,b)>}{T_{n}(a,b)}} {\sum^{\langle k\rangle -1}_{m=0}\frac{\alpha_{m}(a,b)<T(a,b)>}{T_{m}(a,b)}}= \frac{\frac{\alpha_{n}(a,b)}{T_{n}(a,b)}} {\sum^{\langle k\rangle -1}_{m=0}\frac{\alpha_{m}(a,b)}{T_{m}(a,b)}}.
\end{equation} $Q_{n}(a,b)$ actually represents the probability that the recipient owns $n$ ($n=0, 1, 2, 3$) infected neighbors expect for the preselected informed one, involving both the spatial and time correlations of the message diffusion.

Herein, we select different parameter regions containing the numerical critical boundaries for corresponding lattices with $n_{s}=2$ for the calculation of $Q_{n}(a,b)$. We average all non-zero $Q_{n}(a,b)$ in the selected parameter ranges for the expected indices $Q_{n}$, and substitute them into the equation
\begin{eqnarray}
T_{C} = \left\langle T\right\rangle=\sum^{\langle k\rangle-1}_{n=0}Q_{n}T_{n}(a,b)
\end{eqnarray} to get the verification thresholds.

\section*{Acknowledgments}
This work was supported by the National Natural Science Foundation of China under Grants No. 11135001 and No. 11105025, and by the Fundamental Research Funds for the Central Universities under Grant No. lzujbky-2014-28.


\newpage 
\large{\textbf{Supplementary Information:}}
\normalsize
\section{Message diffusion on square lattice}
\label{square}

We first present additional results to support the arguments for message diffusion on square lattice.

In Fig.~S\ref{lattice1470}, we can observe that the size of message spreading  depends more on the stickiness of the message (values of $a$ with fixed $n_{s}$) than on the persistence $b$. We observe that the information can reach the vast majority of population (more than $80\%$) for $a\gtrsim 0.45$.

Fig.~S\ref{laalpha} shows the dependence of the size of recovered population on the parameters $a$ and $b$. For $a\lesssim0.3$, $\alpha_{0}\approx 0.7$, whose value is much larger than the other three indices $\alpha_{n}$ $(n=1,2,3)$, which indicates that the information has not yet outbreak. As $a$ increases, transmission events $E_{m}$ ($m>0$) contribute a lot to the message spreading, hinting the large scale outbreak of the message. Accordingly, $\alpha_{0}$ ($\alpha_{n}$) decreases (increases) sharply, as shown in Figs.~S\ref{laalpha}(b)--(d). In addition, it is found that $E_{0}$ and $E_{1}$ (depending more closely on $a$) constitute most of the transmission events, whereas $E_{2}$ and $E_{3}$ rarely occurs during the spreading process.

To provide support for the real phenomena ``three men make a tiger" (or ``A lie, if repeated often enough, will be accepted as truth"), in what follows we investigate the changes of the accumulative indices $\eta_{i}(a,b)$ by evaluating verification indices $\alpha_{i}$ as $\eta_{i}=\sum_{j<i}\alpha_{j}$ ($i=1,2,3,4$). Here the accumulative indices $\eta_{i}$ represents the proportions of individuals that have adopted the message when they heard it from at most $i$ informed neighbors. Like the results in Fig.~S\ref{laalpha}, high values of $\eta_{1}$  reflect that the occurrences of $E_{0}$ and $E_{1}$ account for most of the transmission events. In addition, the results in Figs.~S\ref{lbar2}(b)(c) show that the vast majority will accept a message as truth if it is mentioned or reported by at least two or three neighbors, which supports the mechanisms ``three men make a tiger" (or ``A lie, if repeated often enough, will be accepted as truth"). Moreover, it also indicates that the reinforcement begins to work as the information is bursting and prevailing on the square lattice, where the growth of $\eta_{i}$ keep increasing with $n_{i}$ (Fig.~S\ref{lbar2}(b)). To make the point clear, we plot the global graphs of the four accumulative indices $\eta_{i}(a,b)$ for different values of inflection point ($n_{s}$) in Fig.~S\ref{lasatu2}, Fig.~S\ref{lasatu3}, and Fig.~S\ref{lasatu4}, respectively. Similar behaviors can be detected in the three figures. The results, especially for positive persistence, also reveal the more important roles of $E_{0}$ and $E_{1}$, and simultaneously provide the evidence for the real phenomenon ``three men make a tiger" (or ``A lie, if repeated often enough, will be accepted as truth").

\section{Message diffusion on random regular networks (RRNs) and regular lattice networks (RLs)}
\label{regular}
In this section, we present additional results to support the arguments for message diffusion on RRNs and RLs.

The sizes of message diffusion are presented in Fig.~S\ref{rhinfor} as a function of $a$ and $b$. In the presence of social reinforcement ($b>0$), we can observe that the message can more easily invade and reach the majority of the population in larger parameter regions beyond the thresholds on RLs, by comparing with that on RRNs with the same average degree. The denser regular lattices (Moore lattice in Fig.~S\ref{rhinfor}(b)~(d)) are much more efficient in promoting information spreading~\cite{science} when the message outbreaks. The reason is that there are more local clustering links can be used for transmission with smaller $n_{s}$~\cite{science,zhou,report} and positive persistence. Instead, the message can more easily diffuse in the RRNs in the presence of strong decay effects ($b<0$). 

The results summarized in Fig.~S\ref{r6alpharound} and Fig.~S\ref{r8alpharound} show that the peaks of $E_{i}(t)$ ($i=0, 1, 2, 3$) brings about the peaks of subsequent transmission event $E_{i+1}(t)$, which indicates that transmission events $E_{i}(t)$ with $i>2$ fail to last stably simultaneously, even at the critical points throughout the spreading processes. There are thus no time correlations among different events. The time correlations of different transmission events can be completely ruled out in estimating the critical behaviors of the message spreading.

By comparison, the occurrences of transmission events in Hexagonal lattices and Moore lattices can last stably for a long time at critical points (see Fig.~S\ref{halpharound2}(b) and Fig.~S\ref{malpharound2}(b)). This suggests the existence of the time correlations among different transmission events $E_{i}(t)$. Besides, the observed huge disparities between $\alpha_{i}(a,b,t)$ and the corresponding $\beta_{i}(a,b,t)$ illustrate that the time correlations among transmission events in the both lattices are considerable.

As observed in Fig.~S\ref{hbar2}(b)(c), Fig.~S\ref{mbar2}(b)(c), and the figures ranging from Fig.~S\ref{hsatu2} to Fig.~S\ref{msatu5}, the message captures the vast majority of the population until $E_{i}(t)$ ($i>2$) happens when message outbreaks and prevails. And the spreading reaches a saturation state for the case where $\frac{n_{s}}{\left\langle k\right\rangle}>\frac{1}{2}$, which means that the transmission events $E_{i}$ ($i>\frac{1}{2}\left\langle k\right\rangle$) rarely happen in the spreading process. Therefore, the results in Fig.~S\ref{hbar2}--Fig.~S\ref{msatu5} can be regarded as the evidence of the mechanism ``Three men make a tiger" (or ``A lie, if repeated often enough, will be accepted as truth"), in addition to the results for positive persistence illustrated in Fig.~S\ref{r6satu2}, and Fig.~S\ref{r8satu2}.

\section{Message diffusion on ER and SF networks}
In this section, we provide additional results to support the arguments for message diffusion on ER networks and SF networks.

We observe in Fig.~S\ref{baalpharound} that transmission events $E_{i}(t)$ ($i>2$) fail to last simultaneously and stably over the long time at the critical points throughout the spreading process. Therefore, the time correlations need not to be taken into consideration in theoretical analysis. We have found the same phenomena in the ER networks and the SF with other average degrees.

As illustrated in Fig.~S\ref{ae8thre}, the theoretical estimations are sufficient to give fairly precise value of the thresholds. It is apparent that the critical behaviors of the message spreading are completely determined by stickiness ($a$) of message. Moreover, in comparison with the case of ER networks, the analytical solutions are in better agreement with the simulation in SF networks, attributing to shorter shortest paths and hubs~\cite{path1,bapath,shortest}. More interestingly, both the analytical boundaries and the numerical thresholds are shifting to left with average degree $\left\langle k\right\rangle$, instead of size of the populations as stressed in~\cite{hub1}. This demonstrates that message can easily reach and infect a larger amount of susceptible individuals through paths from those high-degree vertices whose links increase rapidly with $\left\langle k\right\rangle$, although they can be infected only once.

In Fig.~S\ref{beinfor}, we find that the persistence also  boosts its reasonable impact on the spreading as $\left\langle k\right\rangle$ gets larger.  It implies that hub nodes of larger size and shorter shortest paths in the networks can transmit the information more efficiently~\cite{path1,hub1,shortest}. It is in accordance with the conclusion in~\cite{degree} that higher degrees and densities are relevant factors in improving the global spread of information.

It should be noted that the both ER and SF networks with small average degrees (such as $\langle k \rangle =6, 8$) are tree-like, with few short loops, indicating that the critical transmissibility $T_{C}$ can be derived from equation~$T_{c} = \frac{\left\langle k\right\rangle}{\left\langle k^{2}\right\rangle -\left\langle k\right\rangle}$~\cite{degree,ther1}. More specifically, $T_{c} = \frac{1}{\left\langle k\right\rangle -1}$ for ER networks. Qualitatively, the contributions of the infection events $E_{n}~(n>\left\langle k\right\rangle-2)$ to the spread are insignificant (see Fig.~S\ref{er6satu} and Fig.~S\ref{ba6satu}). Therefore we neglect the contributions of the transmission events $E_{l}(t)$ ($l>\langle k\rangle$) to message spreading, and further assume that the following relationship $T_{C} = \left\langle T\right\rangle = \sum^{\left\langle k\right\rangle-1}_{n=0}q_{n}(1-e^{-\lambda_{n+1}(a,b)})$ is always satisfied in estimating the analytical thresholds. 

Similar to what has been demonstrated in Sec.~\ref{regular}, both Fig.~S\ref{er6satu} (ER networks) and Fig.~S\ref{ba6satu} (SF networks) show that the results nearby the critical points for $b>0$ can also be considered as the evidence of the real phenomenon ``Three men make a tiger" (or ``A lie, if repeated often enough, will be accepted as truth").

\newpage

\setcounter{figure}{0}
\renewcommand\thefigure{S\arabic{figure}}
\section*{\normalsize{Figures}}
\label{figures}
\begin{figure}[ht]
\centering
\includegraphics[width=\textwidth]{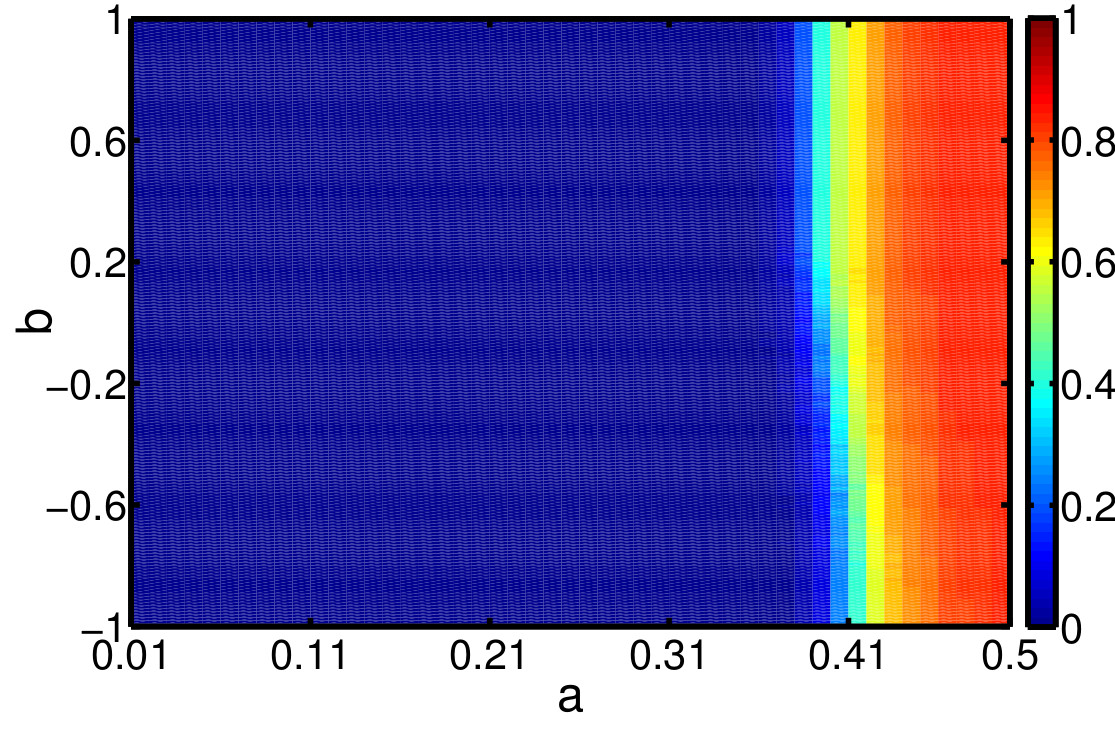}
\caption{\textbf{The densities of recovered individuals as a function of $a$ and $b$.} Each data point is obtained by averaging $100$ independent realizations. The other parameters are $n_{s}=2$ and  $L=101$.}
\label{lattice1470}
\end{figure}

\begin{figure}[ht]
\centering
\includegraphics[width=\textwidth]{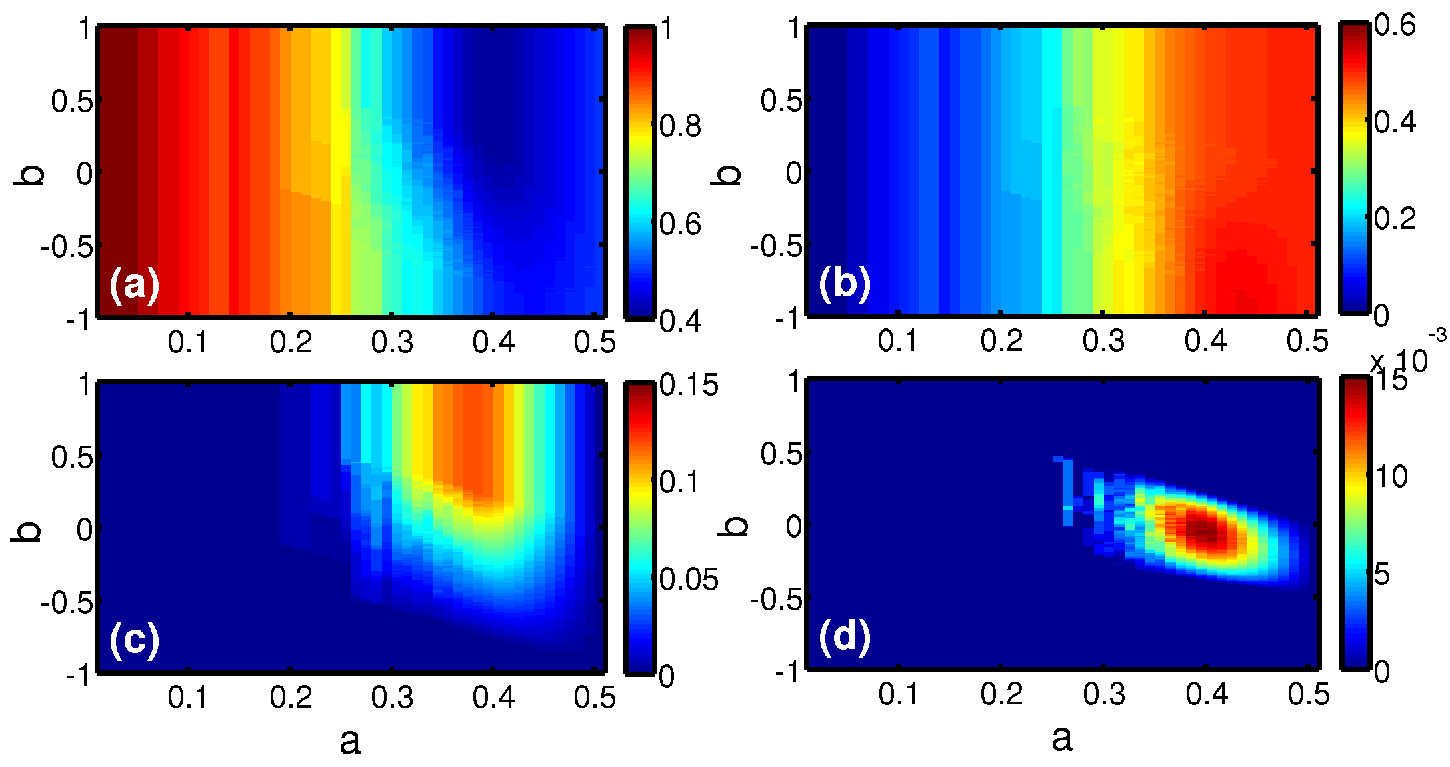}
\caption{\textbf{The four indices (a) $\alpha_{0}(a,b)$, (b) $\alpha_{1}(a,b)$, (c) $\alpha_{2}(a,b)$, and (d) $\alpha_{3}(a,b)$ as a function of $a$ and $b$.} The other parameters are token as $n_{s}=2$ and $L=101$. Each data point is obtained by averaging $100$ independent realizations. As message outbreaks ($a\gtrsim 0.32$), it is clear in (a) and (b) that the transmission events $E_{0}$ and $E_{1}$ contribute the most to the whole spreading process.
}
\label{laalpha}
\end{figure}

\begin{figure}[ht]
\centering
\includegraphics[width=\textwidth]{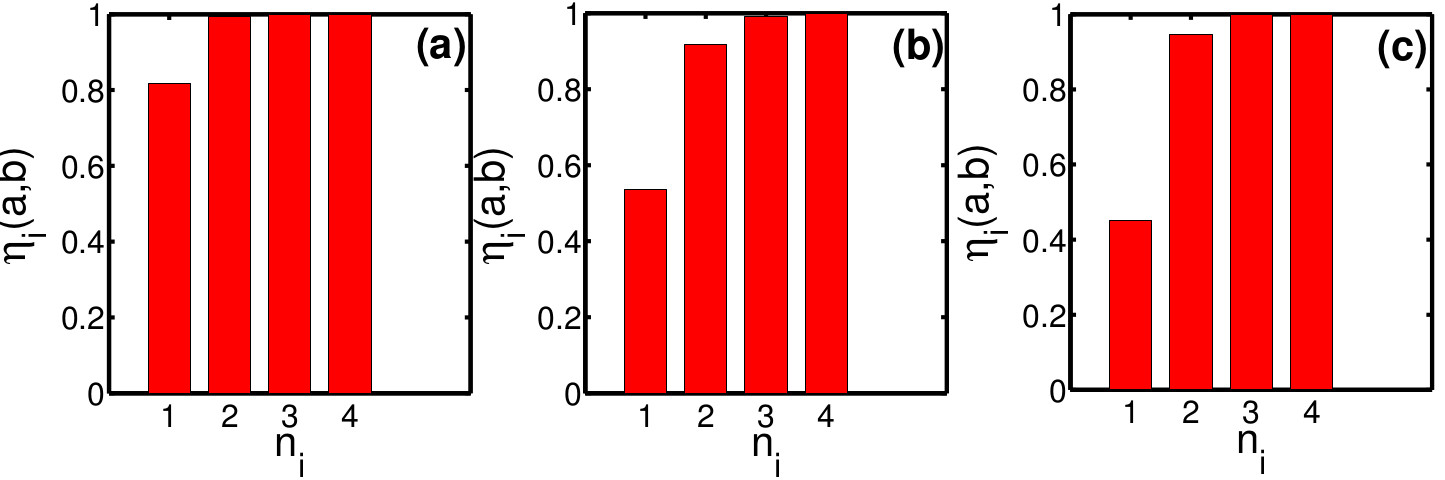}
\caption{\textbf{The four accumulative indices $\eta_{i}(a,b)$ in square latttice.} $n_{i}$ denotes the number of informed neighbors an individual has had at most when it approves the message. Analysis is performed at (a) subcritical point $a=0.20$, $b=0.20$; (b) critical point $a=0.33$, $b=0.20$; and (c) supercritical point $a=0.45$, $b=0.20$. The other parameters are token as $n_{s}=2$ and $L=101$. Results are obtained by averaging $100$ independent realizations. No matter which case, the diffusion of the message owes much to $E_{0}(t)$ and $E_{1}(t)$. In (b) and (c), the value of $\eta_{3}$ approaches to $1$ rather than $\eta_{i}$ ($i<4$) and $\eta_{i}$ ($i<3$),  and the gaps between $\eta_{1}$ and $\eta_{2}$ are obvious. The results in (b) and (c) can be considered as an evidence of the emergence of ``Three men make a tiger" (or ``A lie, if repeated often enough, will be accepted as truth") -- the vast majority of the population accept the message as truth only when it is repeated more than two times in their ears.}
\label{lbar2}
\end{figure}

\begin{figure}[ht]
\centering
\includegraphics[width=\textwidth]{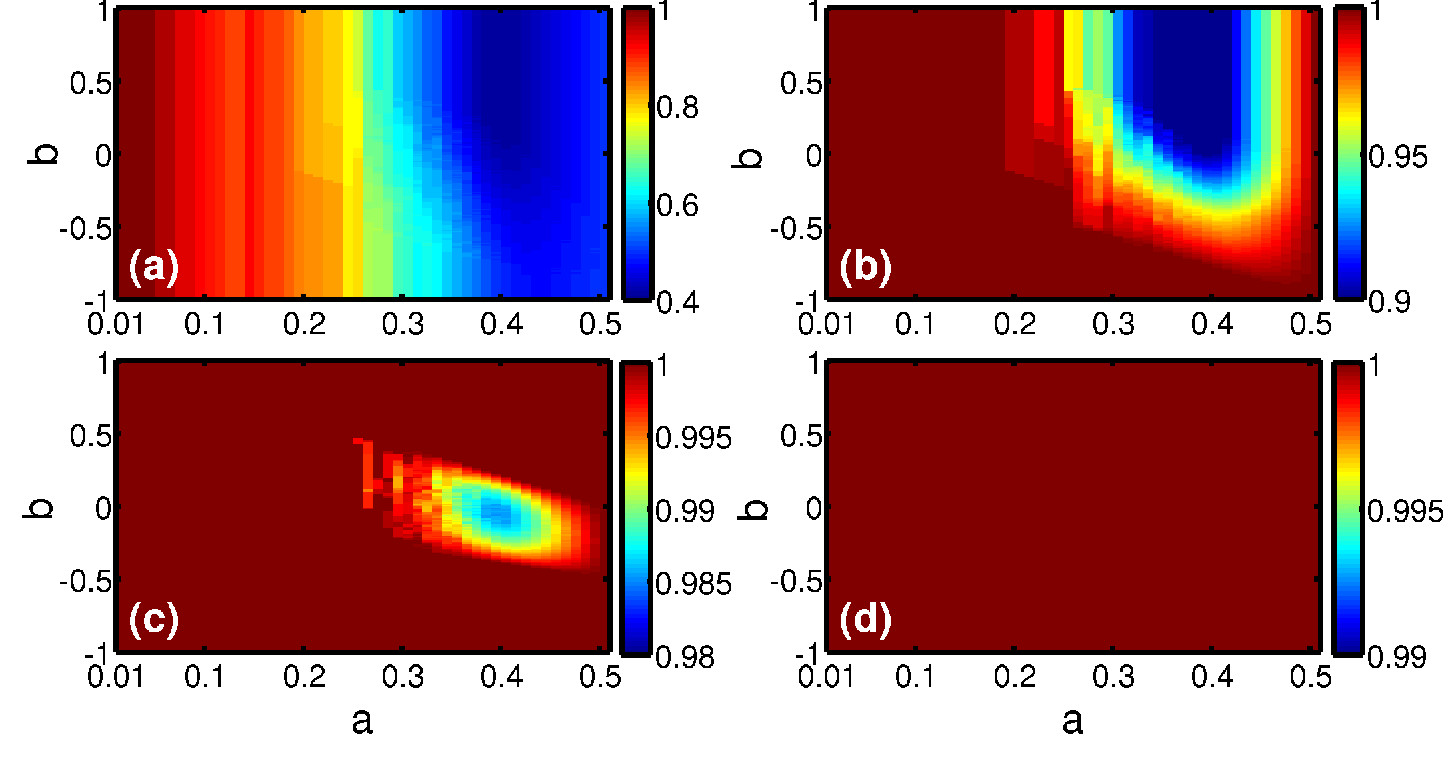}
\caption{\textbf{The four accumulative indices $\eta_{0}$ (a), $\eta_{1}$ (b), $\eta_{2}$ (c), and $\eta_{3}$ (d) as a function of $a$ and $b$.} The other parameters are token as $n_{s}=2$ and $L=101$. Each data point is obtained by averaging $100$ independent realizations. In the wide parameter regions for positive persistence, especially nearby the critical boundary (threshold), only $\eta_{i}>90\%$ ($i\geq 2$) rather than $\eta_{1}$, where transmission events $E_{i}(t)$ ($i\geq 2$) contribute more than $60\%$ of the scale of message spreading (i.e., the spreading thus reaches a saturation level). It also implies that the majority of the population accept the message as truth only if the message is repeated more than three times in their ears. Therefore, the results with positive persistence nearby the threshold also provide the evidence for the real phenomenon ``Three men make a tiger" (or ``A lie, if repeated often enough, will be accepted as truth").}
\label{lasatu2}
\end{figure}

\begin{figure}[ht]
\centering
\includegraphics[width=\textwidth]{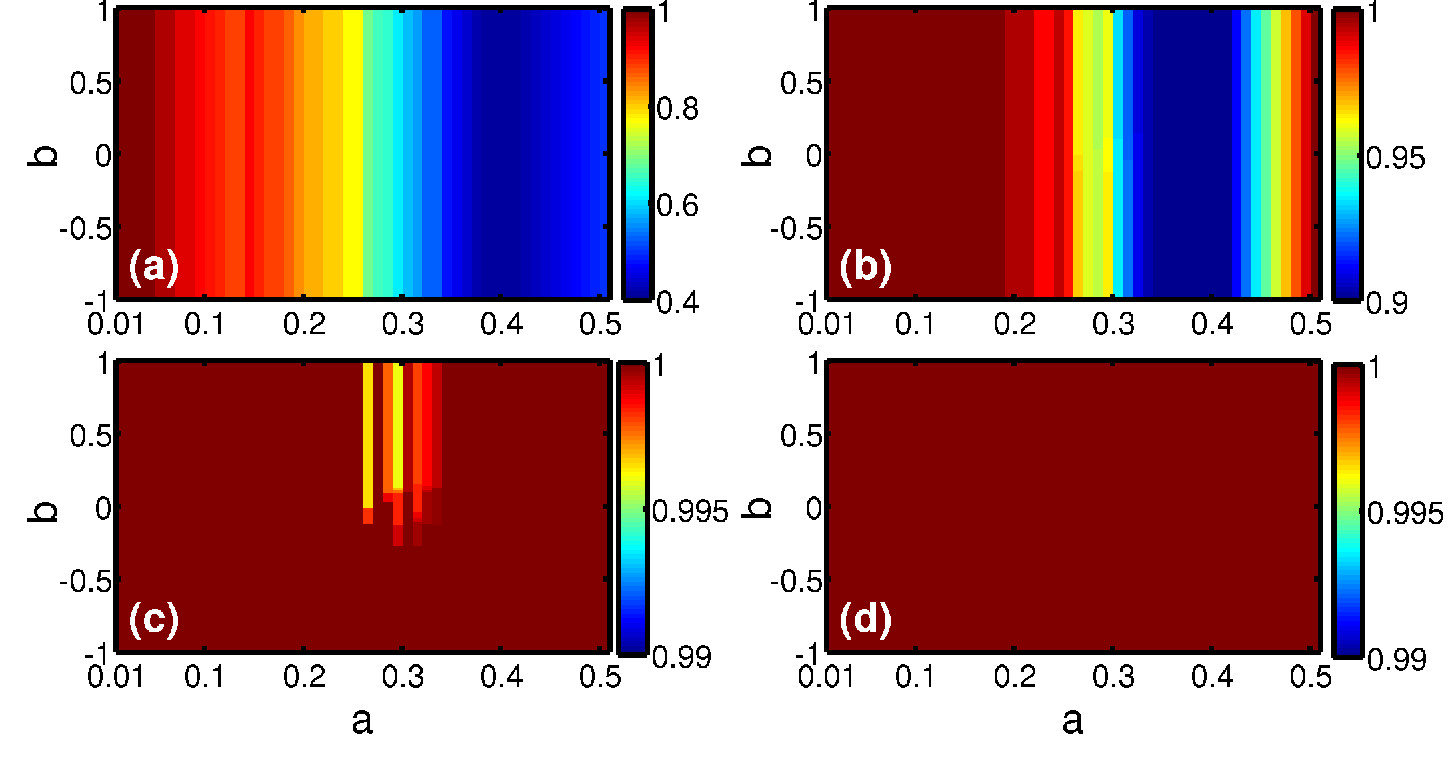}
\caption{\textbf{The four accumulative indices $\eta_{0}$ (a), $\eta_{1}$ (b), $\eta_{2}$ (c), and $\eta_{3}$ (d) as a function of $a$ and $b$.} The other parameters are token as $n_{s}=3$ and $L=101$. Each data point is obtained by averaging $100$ independent realizations. In the wide parameter regions for positive persistence, especially nearby the critical boundary (threshold), only $\eta_{i}>90\%$ ($i\geq 2$) rather than $\eta_{1}$, where transmission events $E_{i}(t)$ ($i\geq 2$) contribute more than $60\%$ of the scale of message spreading (i.e., the spreading thus reaches a saturation level). It also implies that the majority of the population accept the message as truth only if the message is repeated more than three times in their ears. Therefore, the results with positive persistence nearby the threshold also provide the evidence for the real phenomenon ``Three men make a tiger" (or ``A lie, if repeated often enough, will be accepted as truth").}
\label{lasatu3}
\includegraphics[width=\textwidth]{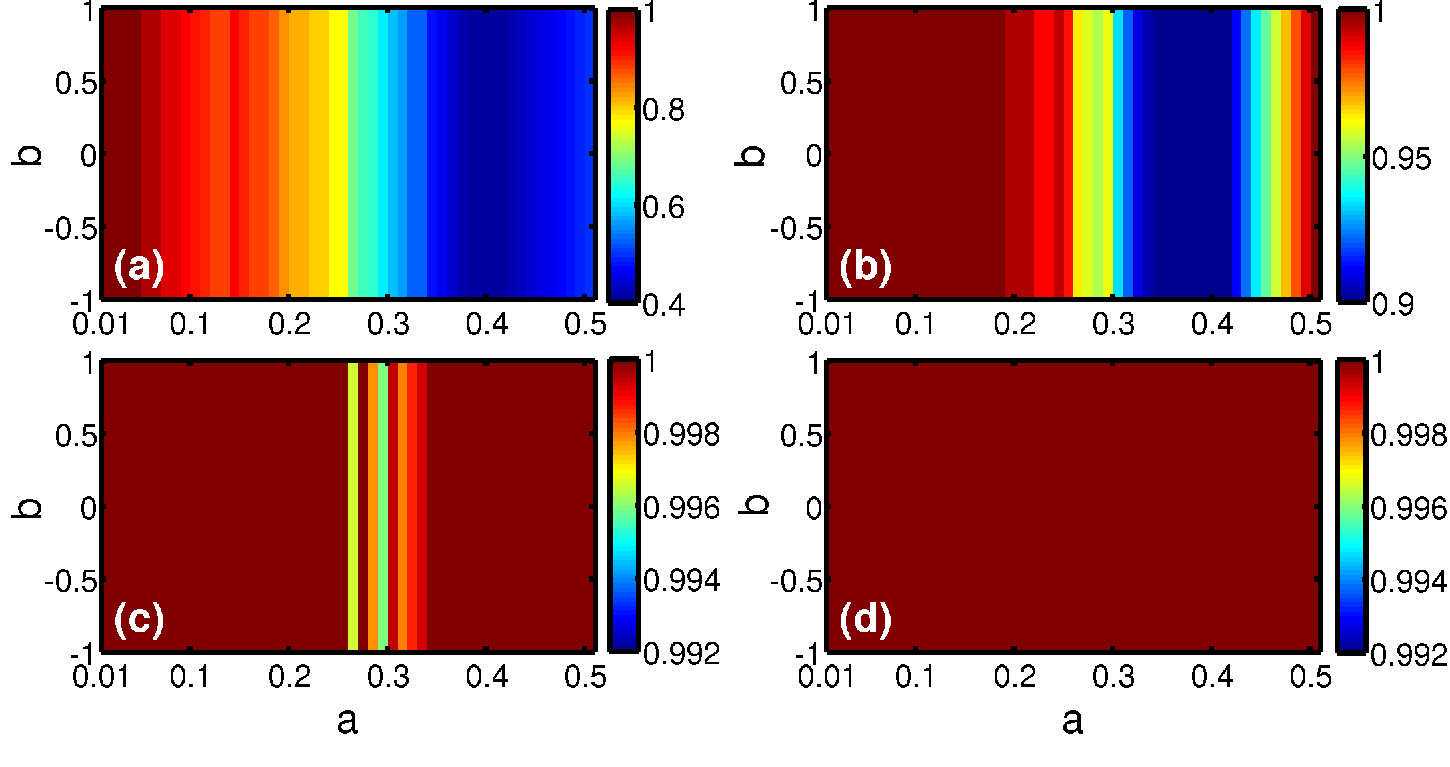}
\caption{\textbf{The four accumulative indices $\eta_{0}$ (a), $\eta_{1}$ (b), $\eta_{2}$ (c), and $\eta_{3}$ (d) as a function of $a$ and $b$.} The other parameters are token as $n_{s}=4$ and $L=101$. Each data point is obtained by averaging $100$ independent realizations. In the wide parameter regions for positive persistence, especially nearby the critical boundary (threshold), only $\eta_{i}>90\%$ ($i\geq 2$) rather than $\eta_{1}$, where transmission events $E_{i}(t)$ ($i\geq 2$) contribute more than $60\%$ of the scale of message spreading (i.e., the spreading thus reaches a saturation level). It also implies that the majority of the population accept the message as truth only if the message is repeated more than three times in their ears. Therefore, the results with positive persistence nearby the threshold also provide the evidence for the real phenomenon ``Three men make a tiger" (or ``A lie, if repeated often enough, will be accepted as truth").}
\label{lasatu4}
\end{figure}

\begin{figure}[ht]
\centering
\includegraphics[width=\textwidth]{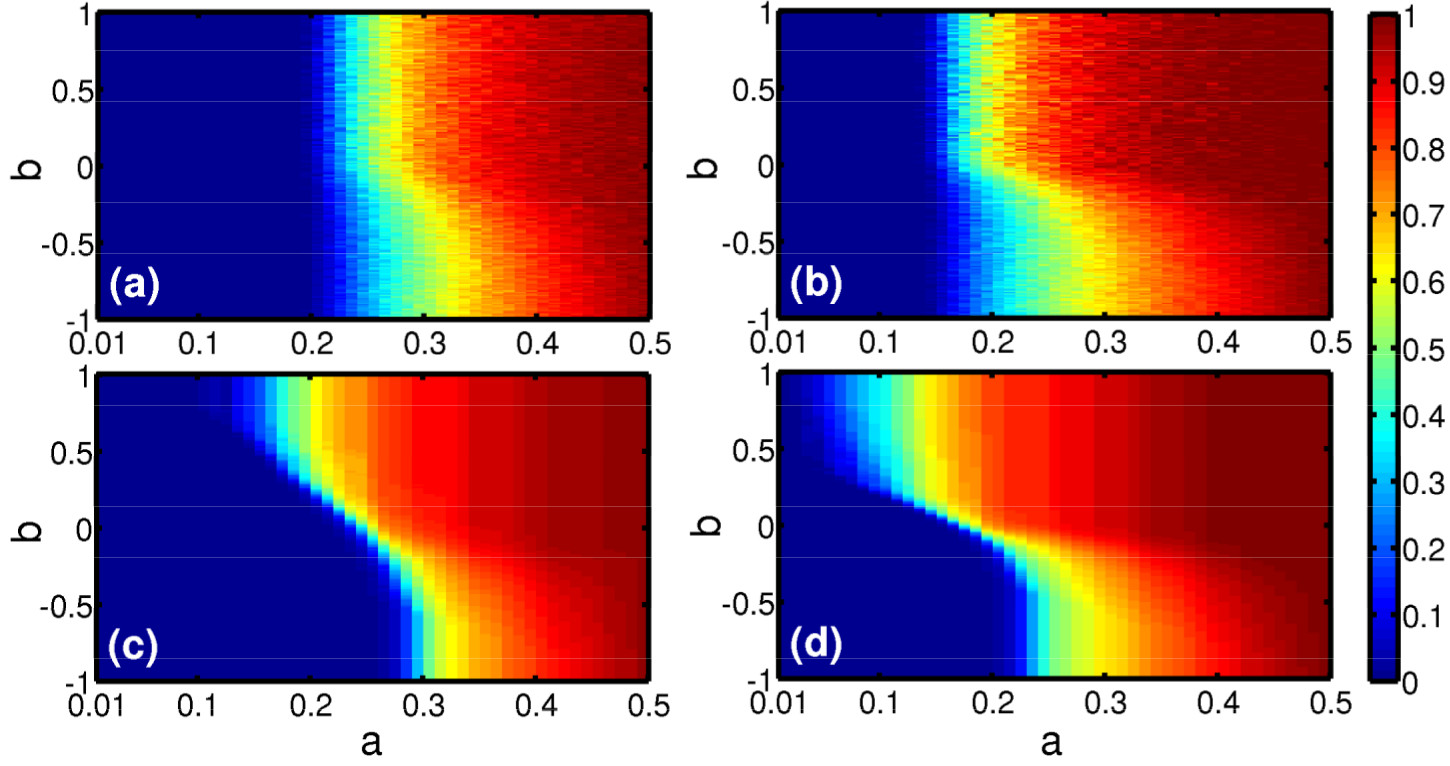}
\caption{\textbf{The densities of recovered individuals as a function of $a$ and $b$.} The other parameters are token as $n_{s}=2$, $N=1000$ for the RRNs, and $L=101$ for the RLs. The scales of spreading on RLs ((a)~(b)) are compared with that in RRNs ((a)~(b)). The degrees of the networks are $<k>=6$ (a, c) and $<k>=8$ (b, d), respectively. Each data is obtained by averaging $100$ independent realizations. By comparing (a) with (c), for $b>0$, it is clear in (c) that the message can more easily outbreak and capture a larger population, in contrast to what is observed in (a), owing to the function of social reinforcement effects. However, more individuals accept the message in RRNs for $b<0$, indicating the advantage of the RRNs in facilitating the diffusion of message in the presence of strong decay effects. As expected, the same conclusion can also be reached by comparing the plots in (b) with that in (d).
}
\label{rhinfor}
\end{figure}

\begin{figure}[ht]
\centering
\includegraphics[width=\textwidth]{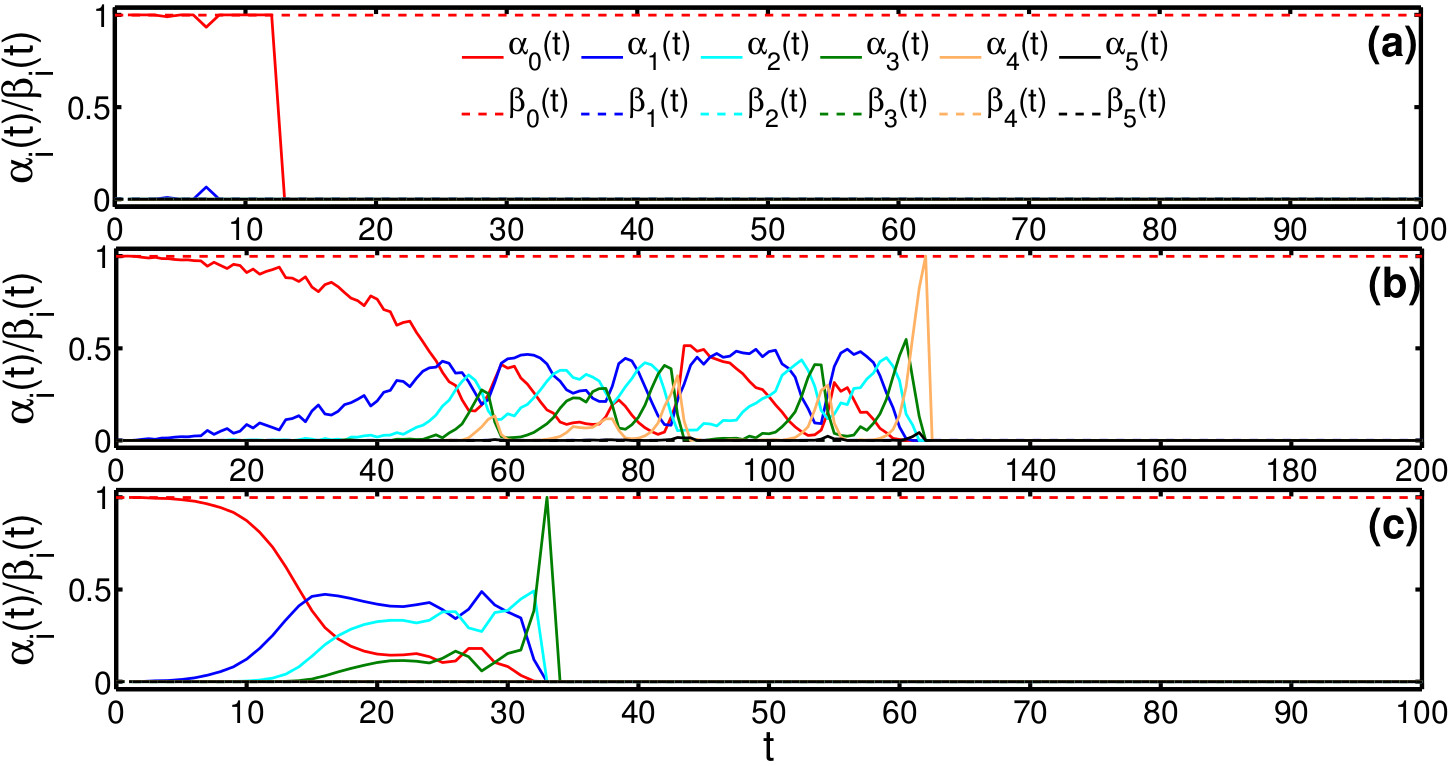}
\caption{\textbf{The evolution of proportions of the transmission events.} The evolution of indices $\alpha_{i}(a,b,t)$ from simulation (solid lines) and $\beta_{i}(a,b,t)$ from prediction of percolation theory (dashed lines) on the RRN with $\langle k\rangle=6$ are presented. Three different cases are considered here: (a) the information vanishes for $a=0.10$, $b=0.20$, (b) it outbreaks for $a=0.19$, $b=0.20$;  and prevails for $a=0.30$, $b=0.20$ (c). The other parameters are token as $n_{s}=2$ and $N=1000$.   It can be observed that the occurrences of all the transmission events fail to last stably. The time correlations among the transmission events can thus be neglected in estimating the critical behavior of the message spreading.}
\label{r6alpharound}
\includegraphics[width=\textwidth]{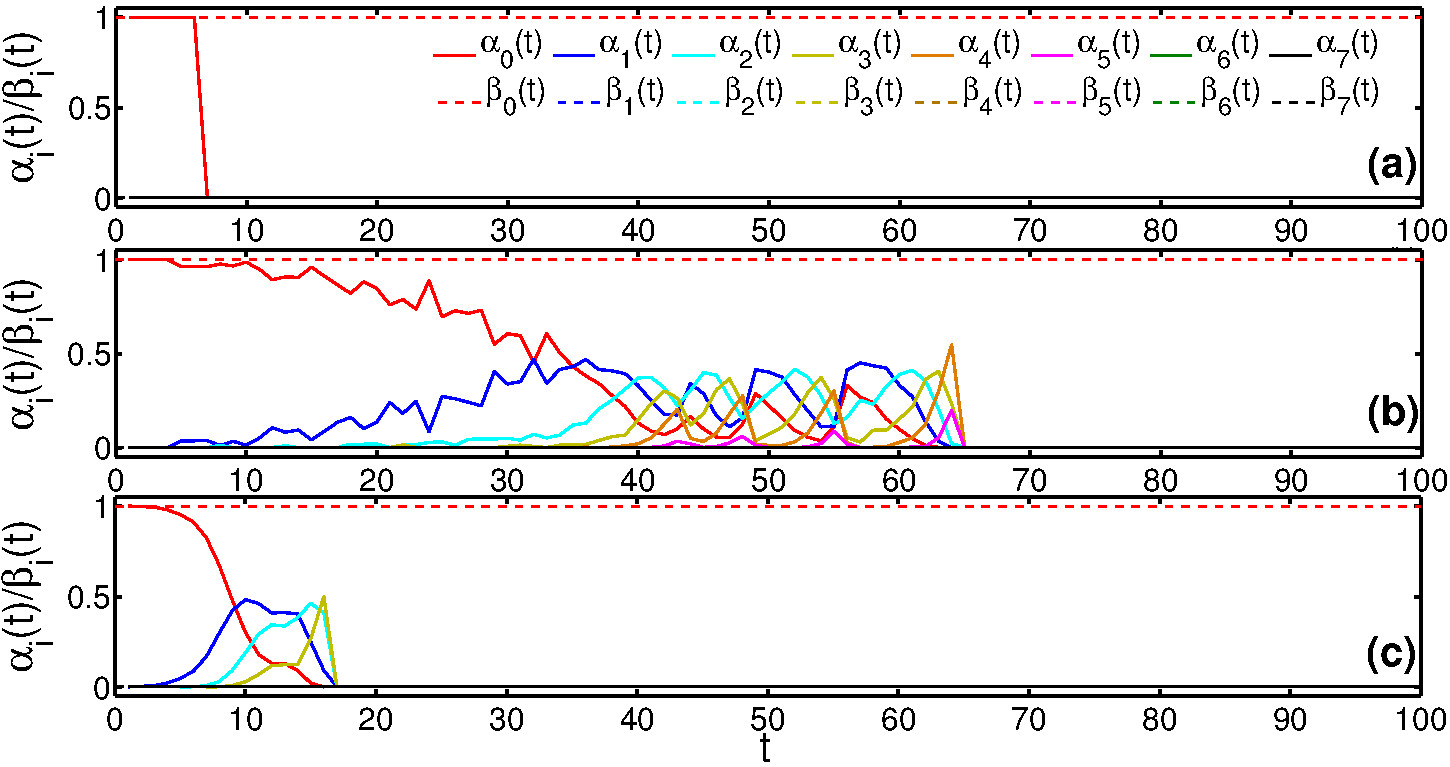}
\caption{\textbf{The evolution of proportions of the transmission events.} The evolution of indices $\alpha_{i}(a,b,t)$ from simulation (solid lines) and $\beta_{i}(a,b,t)$ from prediction of percolation theory (dashed lines) on the RRN with $\langle k\rangle=8$ are presented. Three different cases are considered here: (a) the information vanishes for $a=0.10$, $b=0.20$, (b) it outbreaks for $a=0.13$, $b=0.20$;  and prevails for $a=0.30$, $b=0.20$ (c).The other parameters are token as $n_{s}=2$ and $N=1000$.  It can be observed that the occurrences of all the transmission events fail to last stably throughout the whole spreading process. The time correlations among the transmission events can thus be neglected in estimating the critical behavior of the message spreading.}
\label{r8alpharound}
\end{figure}

\begin{figure}[ht]
\centering
\includegraphics[width=\textwidth]{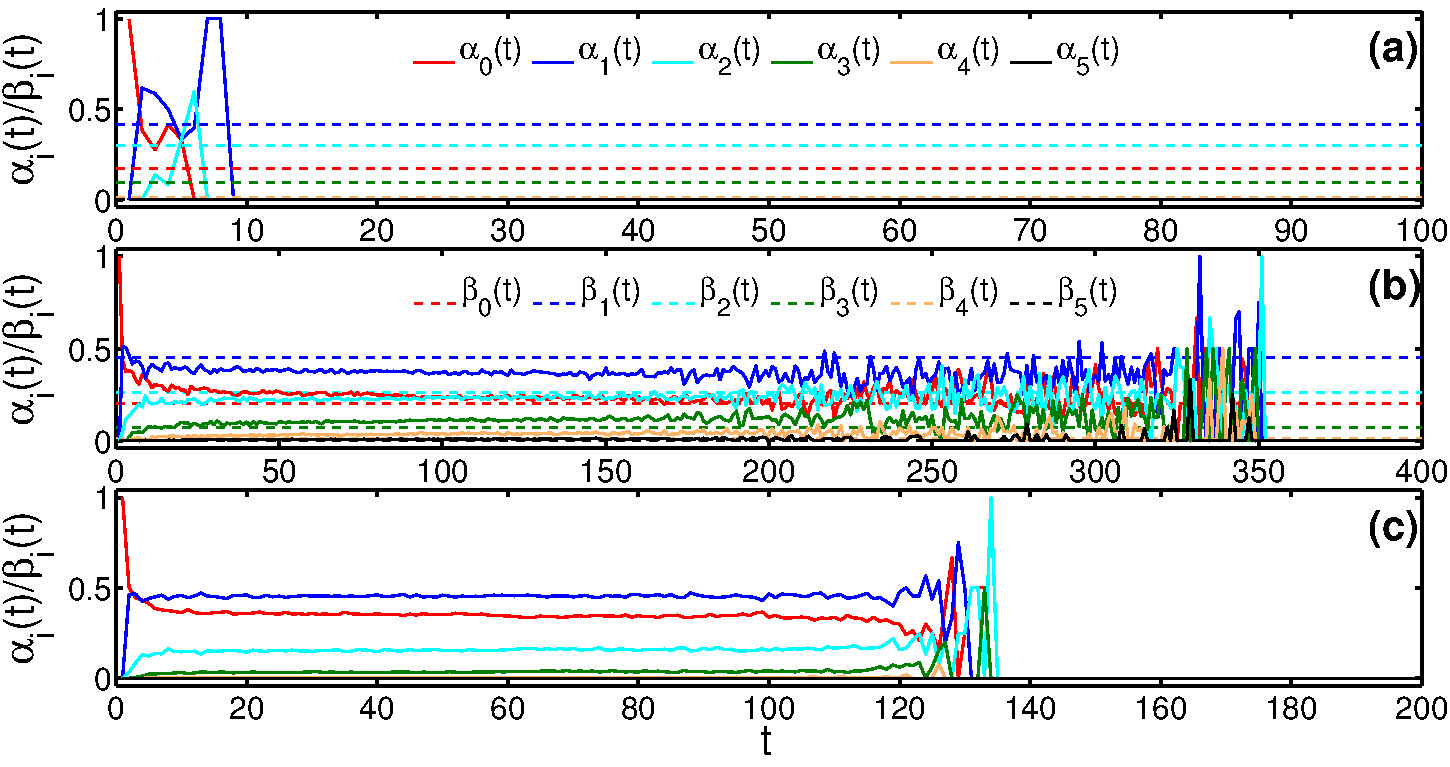}
\caption{\textbf{The evolution of proportions of the transmission events.} The evolution of indices $\alpha_{i}(a,b,t)$ from simulation (solid lines) and $\beta_{i}(a,b,t)$ from prediction of percolation theory (dashed lines) on the Hexagonal lattice are presented. Three difference cases are considered here: (a) the information vanishes for $a=0.10$, $b=0.20$; (b) it outbreaks for $a=0.19$, $b=0.20$; and prevails for $a=0.30$, $b=0.20$ (c). The other parameters are token as $n_{s}=2$ and $L=101$. In comparison with the dynamic behaviors of the corresponding RRNs shown in Fig.~S\ref{r6alpharound}, the occurrences of transmission events in Hexagonal lattice can last stably for over $350$ MCs at the critical point, hence the time correlations among the events cannot be neglected in estimating the critical behavior of the message spreading. The inconsistencies between $\alpha_{i}(a,b,t)$ and the corresponding $\beta_{i}(a,b,t)$ have been chosen to demonstrate the existence of the time correlations between different transmission events $E_{i}(t)$. The phenomenon $\alpha_{i}(t)>0$ in (b) indicates that almost all transmission events involve in the spreading process at the critical point. In addition, $\beta_{i}(t)$ gets close to corresponding $\alpha_{i}(t)$ in (b) and (c).}
\label{halpharound2}
\includegraphics[width=\textwidth]{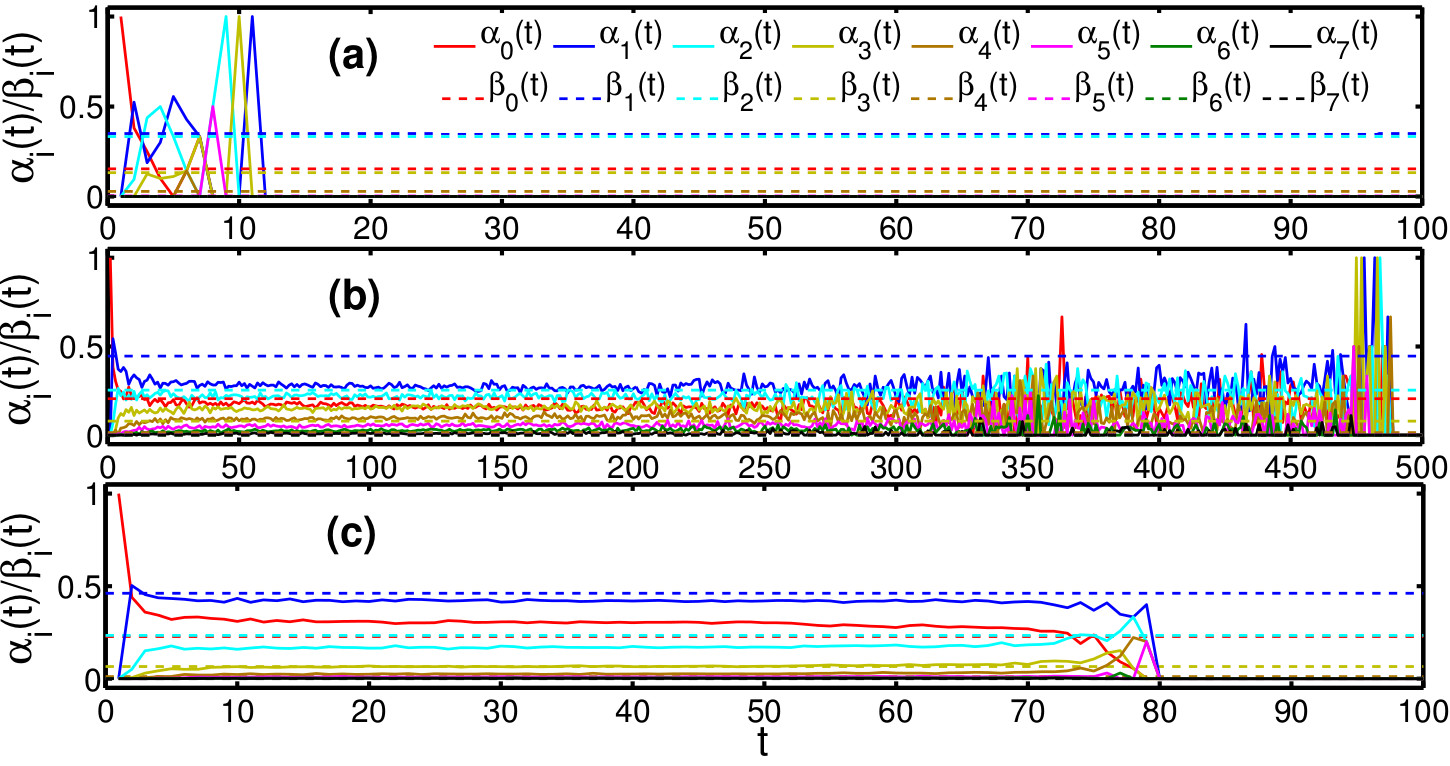}
\caption{\textbf{The evolution of proportions of the transmission events.} The evolution of indices $\alpha_{i}(a,b,t)$ from simulation (solid lines) and $\beta_{i}(a,b,t)$ from prediction of percolation theory (dashed lines) on the Moore lattice are presented. Three different cases are considered here: (a) the information vanishes for $a=0.05$, $b=0.20$; (b) it outbreaks for $a=0.10$, $b=0.20$ and prevails for $a=0.30$, $b=0.20$ (c). The other parameters are token as $n_{s}=2$ and $L=101$. In comparison with the dynamic behaviors of corresponding RRNs shown in Fig.~S\ref{r8alpharound}, the occurrences of transmission events in Moore lattice can last stably for over $450$ MCs at the critical point, hence the time correlation among the events cannot be neglected in estimating the critical behavior of the message spreading. The inconsistencies between $\alpha_{i}(a,b,t)$ and the corresponding $\beta_{i}(a,b,t)$ have been chosen to demonstrate the existence of the time correlations between different transmisssion events $E_{i}(t)$. The phenomenon $\alpha_{i}(t)>0$ in (b) indicates that almost all transmission events involve in the spreading process at the critical point. In addition, $\beta_{i}(t)$ gets close to corresponding $\alpha_{i}(t)$ in (b) and (c).}
\label{malpharound2}
\end{figure}

\begin{figure}[ht]
\centering
\includegraphics[width=\textwidth]{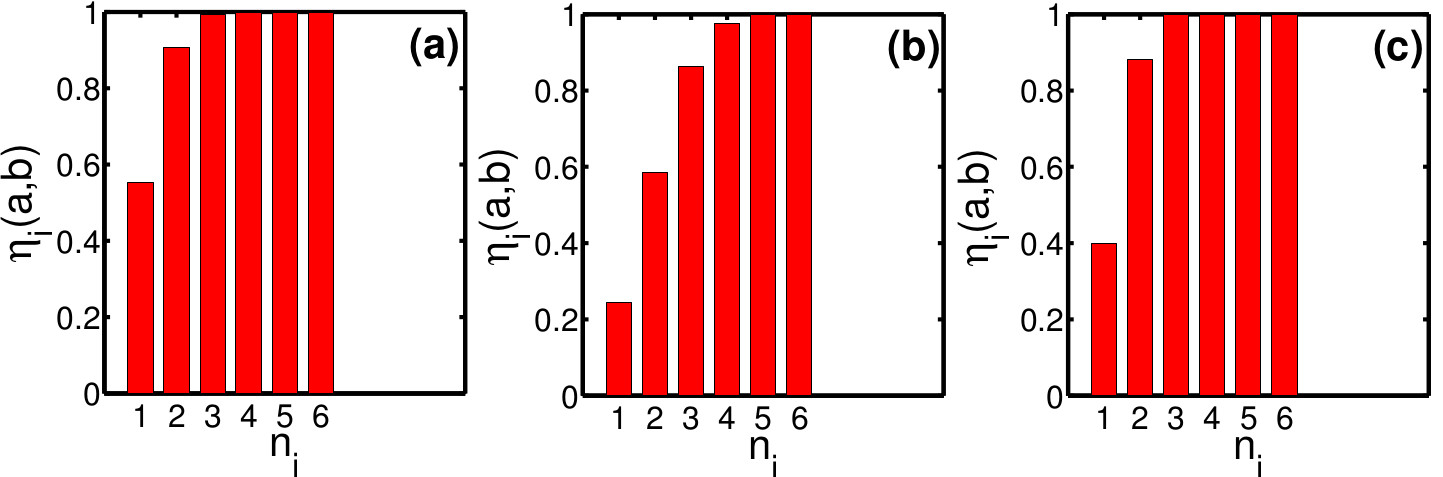}
\caption{\textbf{The six accumulative indices $\eta_{i}(a,b)$ in Hexagonal lattice.} $n_{i}$ denotes the number of informed neighbors an individual has had at most when it approves the message. The parameters are chosen at three selected parameter points: (a) subcritical point $a=0.10$, $b=0.20$; (b) critical point $a=0.18$, $b=0.20$ and (c) supercritical point $a=0.40$, $b=0.20$. The other parameters are token as $n_{s}=2$ and $L=101$. Results are obtained by averaging $100$ independent realizations. In (b), the gaps between $\eta_{i}$ and $\eta_{i+1}$ ($i<4$) are apparent, which indicates that almost all transmission events involve in the spreading process. In (b) and (c), the value of $\eta_{4}$ or $\eta_{3}$ approach to $1$, rather than $\eta_{i}$ ($i<4$) and $\eta_{i}$ ($i<3$). That can be considered as an evidence of the emergence of ``Three men make a tiger" (or ``A lie, if repeated often enough, will be accepted as truth") -- the vast majority of the population accept the message as truth only when it is repeated more than two times in their ears.}
\label{hbar2}
\includegraphics[width=\textwidth]{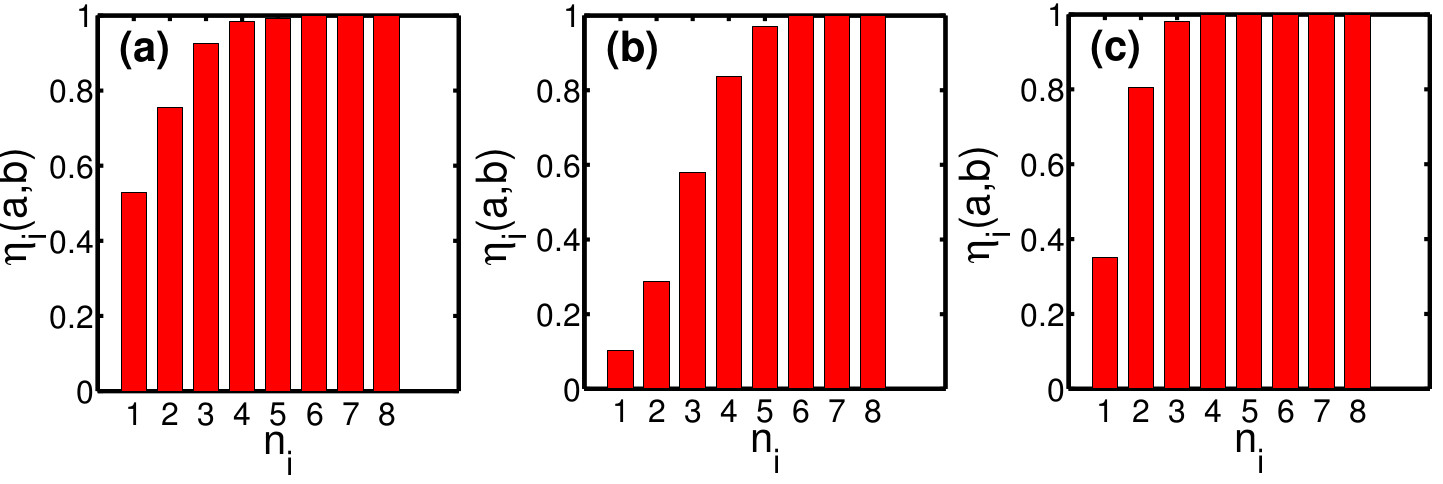}
\caption{\textbf{The eight accumulative indices $\eta_{i}(a,b)$ in Moore lattice.} $n_{i}$ denotes the number of informed neighbors an individual has had at most when it approves the message. The parameters are chosen at three selected parameter points: (a) subcritical point $a=0.10$, $b=0.20$; (b) critical point $a=0.18$, $b=0.20$ and (c) supercritical point $a=0.40$, $b=0.20$. The other parameters are token as $n_{s}=2$ and $L=101$. Results are obtained by averaging $100$ independent realizations. In (b), the gaps between $\eta_{i}$ and $\eta_{i+1}$ ($i<4$) are apparent, which indicates that almost all transmission events involve in the spreading process. In (b) and (c), the value of $\eta_{4}$ or $\eta_{3}$ approach to $1$, rather than $\eta_{i}$ ($i<6$) and $\eta_{i}$ ($i<4$). That can be considered as an evidence of the emergence of ``Three men make a tiger" (or ``A lie, if repeated often enough, will be accepted as truth") -- the vast majority of the population accept the message as truth only when it is repeated more than two times in their ears.}
\label{mbar2}
\end{figure}

\begin{figure}[ht]
\centering
\includegraphics[width=\textwidth]{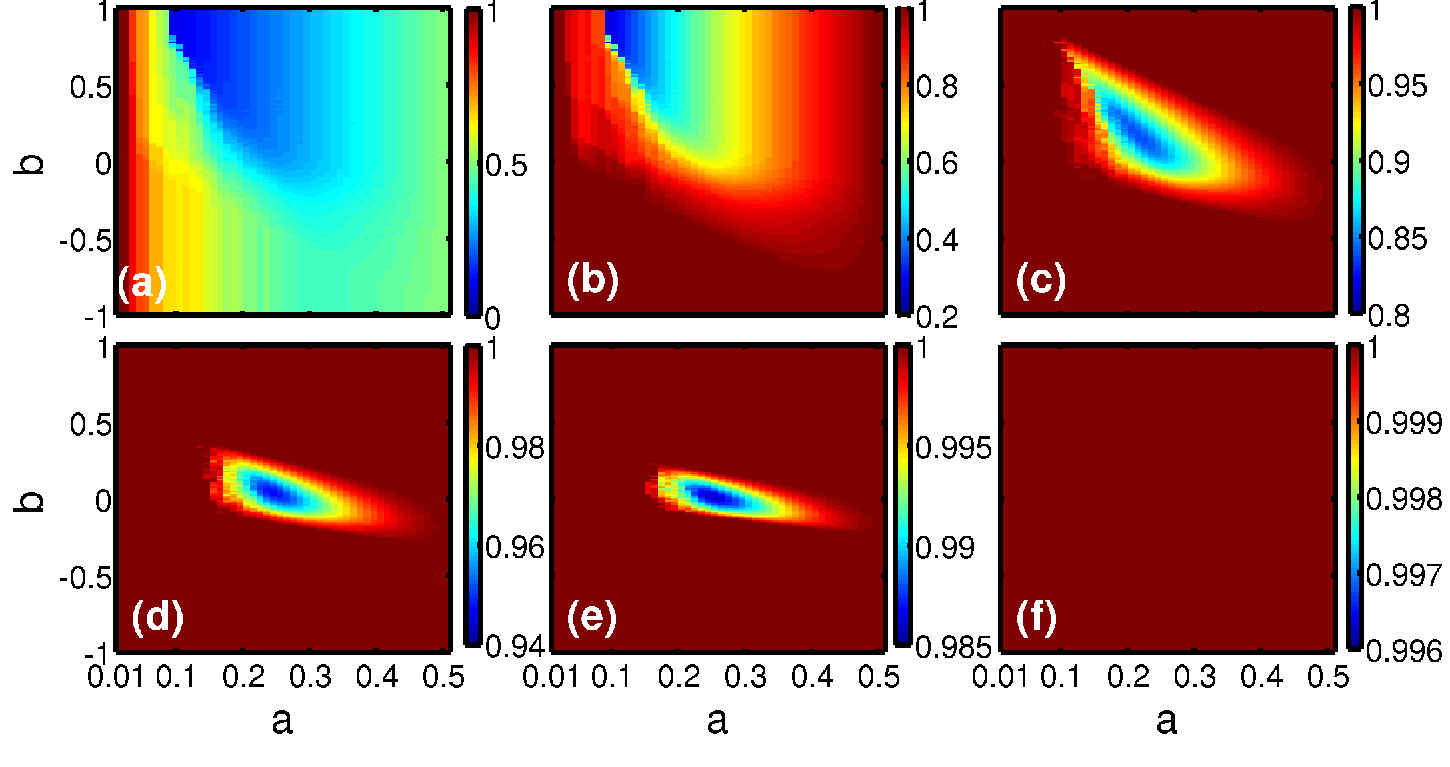}
\caption{\textbf{ The six accumulative indices $\eta_{1}$ (a), $\eta_{2}$ (b), $\eta_{3}$ (c), $\eta_{4}$ (d), $\eta_{5}$ (e), and $\eta_{6}$ (f) as a function of $a$ and $b$ for Hexagonal lattice.} The other parameters are token as $n_{s}=2$ and $L=101$. Each data point is obtained by averaging $100$ independent realizations. In the wide parameter regions for positive persistence, especially nearby the critical boundary (threshold), only $\eta_{i}>80\%$ ($i\geq frac{\langle k \rangle}{2}$) rather than $\eta_{j}$ ($j<3$), where transmission events $E_{i}(t)$ ($i\geq 2$) contribute more than $70\%$ of the scale of message spreading (i.e., the spreading thus reaches a saturation level). It also implies that the majority of the population accept the message as truth only if the message is repeated more than three times in their ears. Therefore, the results with positive persistence nearby the threshold also provide the evidence for the real phenomenon ``Three men make a tiger" (or ``A lie, if repeated often enough, will be accepted as truth").}
\label{hsatu2}
\includegraphics[width=\textwidth]{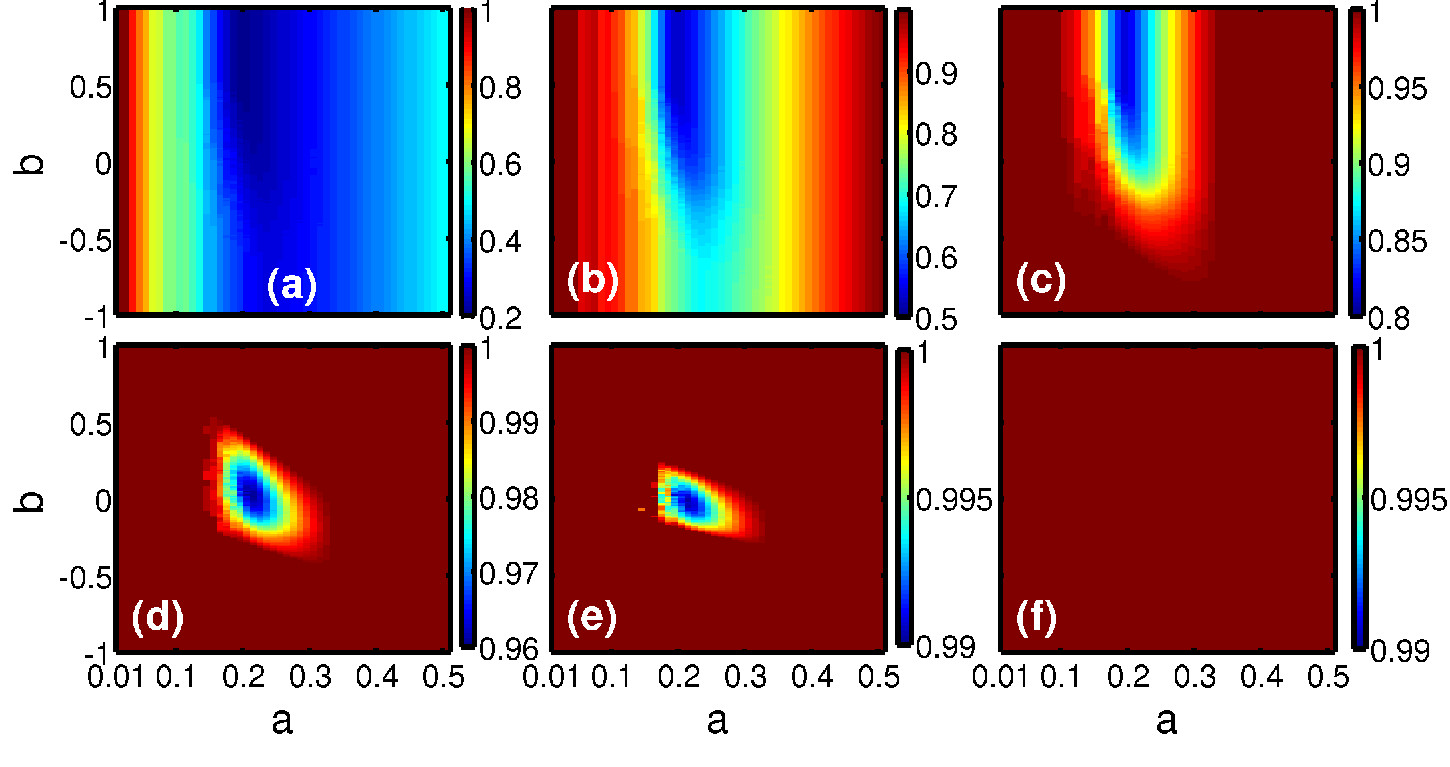}
\caption{\textbf{ The six accumulative indices $\eta_{1}$ (a), $\eta_{2}$ (b), $\eta_{3}$ (c), $\eta_{4}$ (d), $\eta_{5}$ (e), and $\eta_{6}$ (f) as a function of $a$ and $b$ for Hexagonal lattice.} The other parameters are token as $n_{s}=3$ and $L=101$. Each data point is obtained by averaging $100$ independent realizations. In the wide parameter regions with positive persistence, especially nearby the critical boundary (threshold), only $\eta_{i}>80\%$ ($i\geq \frac{\langle k \rangle}{2}$) rather than $\eta_{j}$ ($j<3$), where transmission events $E_{i}(t)$ ($i\geq 2$) contribute more than $80\%$ of the scale of message spreading (i.e., the spreading thus reaches a saturation level). It also implies that the majority of the population accept the message as truth only if the message is repeated more than three times in their ears. Therefore, the results with positive persistence nearby the threshold also provide the evidence for the real phenomenon ``Three men make a tiger" (or ``A lie, if repeated often enough, will be accepted as truth").
}
\label{hsatu3}
\end{figure}

\begin{figure}[ht]
\centering
\includegraphics[width=\textwidth]{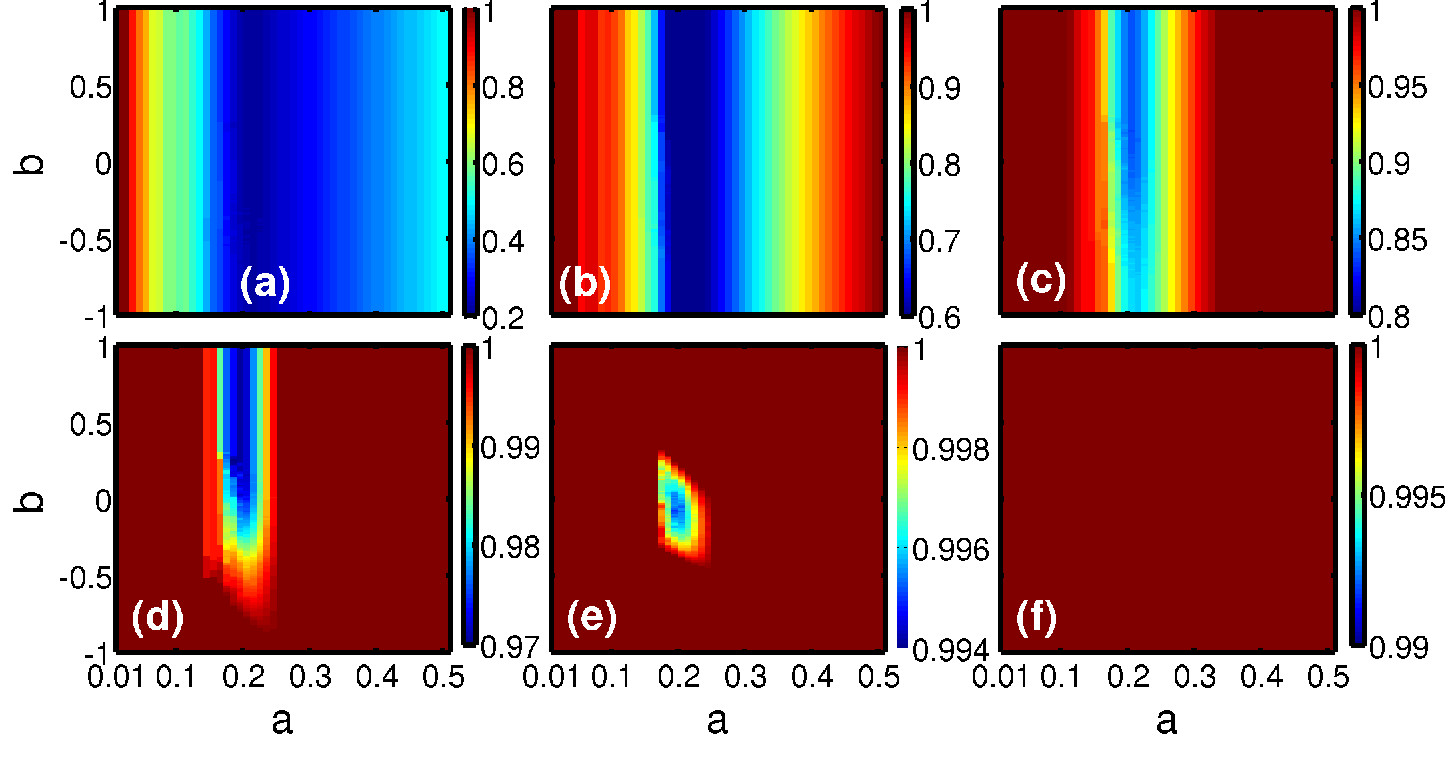}
\caption{\textbf{ The six accumulative indices $\eta_{1}$ (a), $\eta_{2}$ (b), $\eta_{3}$ (c), $\eta_{4}$ (d), $\eta_{5}$ (e), and $\eta_{6}$ (f) as a function of $a$ and $b$ for Hexagonal lattice.} The other parameters are token as $n_{s}=4$ and $L=101$. Each data point is obtained by averaging $100$ independent realizations. In the wide parameter regions for positive persistence, especially nearby the critical boundary (threshold), only $\eta_{i}>80\%$ ($i\geq \frac{\langle k \rangle}{2}$) rather than $\eta_{j}$ ($j<3$), where transmission events $E_{i}(t)$ ($i\geq 2$) contribute more than $80\%$ of the scale of message spreading (i.e., the spreading thus reaches a saturation level). It also implies that the majority of the population accept the message as truth only if the message is repeated more than three times in their ears. Therefore, the results with positive persistence nearby the threshold also provide the evidence for the real phenomenon ``Three men make a tiger" (or ``A lie, if repeated often enough, will be accepted as truth").
}
\label{hsatu4}
\includegraphics[width=\textwidth]{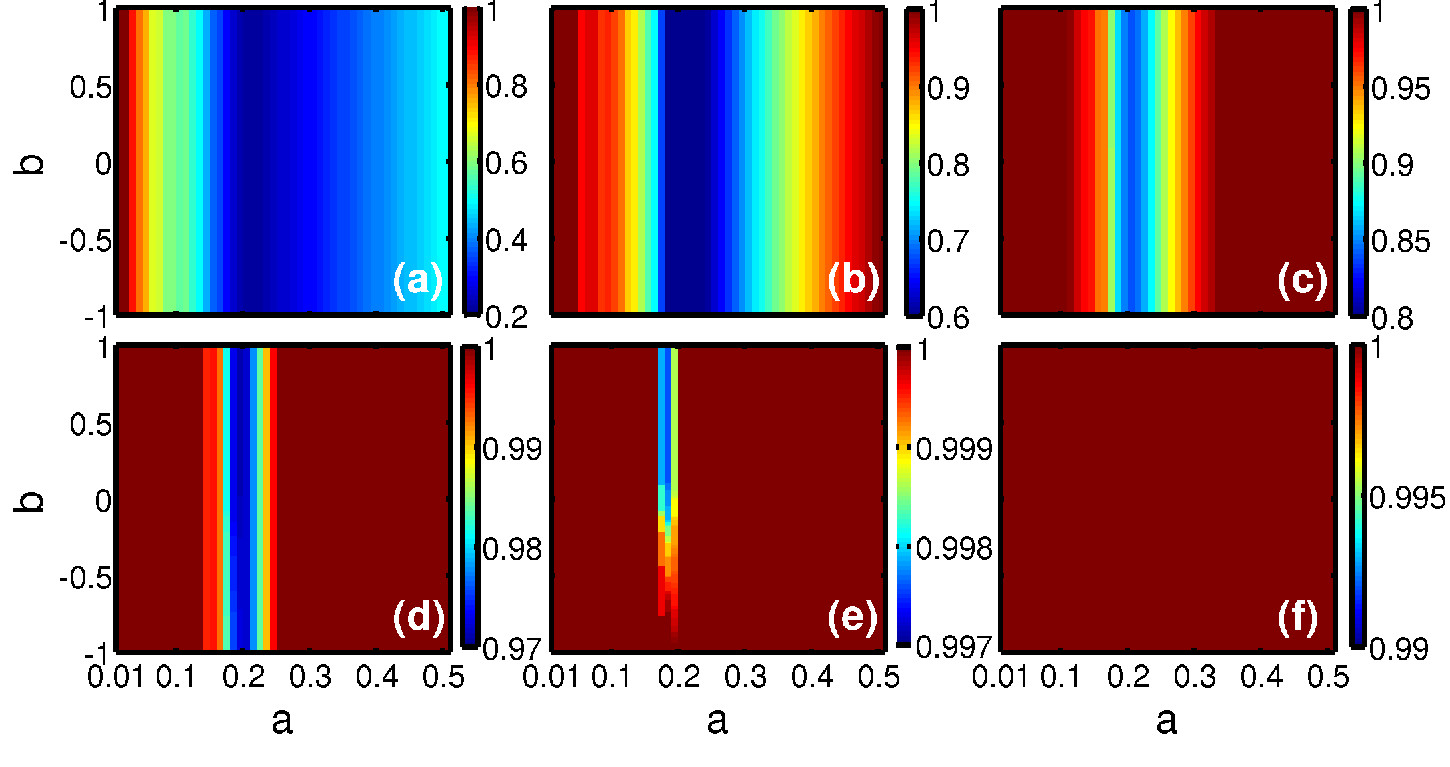}
\caption{\textbf{The six accumulative indices $\eta_{1}$ (a), $\eta_{2}$ (b), $\eta_{3}$ (c), $\eta_{4}$ (d), $\eta_{5}$ (e), and $\eta_{6}$ (f) as a function of $a$ and $b$ for Hexagonal lattice.} The other parameters are token as $n_{s}=5$ and $L=101$. Each data point is obtained by averaging $100$ independent realizations.   In the wide parameter regions, especially nearby the critical boundary (threshold), only $\eta_{i}>80\%$ ($i\geq \frac{\langle k \rangle}{2}$) rather than $\eta_{j}$ ($j<3$), where transmission events $E_{i}(t)$ ($i\geq 2$) contribute more than $80\%$ of the scale of message spreading (i.e., the spreading thus reaches a saturation level). It also implies that the majority of the population accept the message as truth only if the message is repeated more than three times in their ears. Therefore, the results nearby the threshold also provide the evidence for the real phenomenon ``Three men make a tiger" (or ``A lie, if repeated often enough, will be accepted as truth").
}
\label{hsatu5}
\end{figure}

\begin{figure}[ht]
\centering
\includegraphics[width=\textwidth]{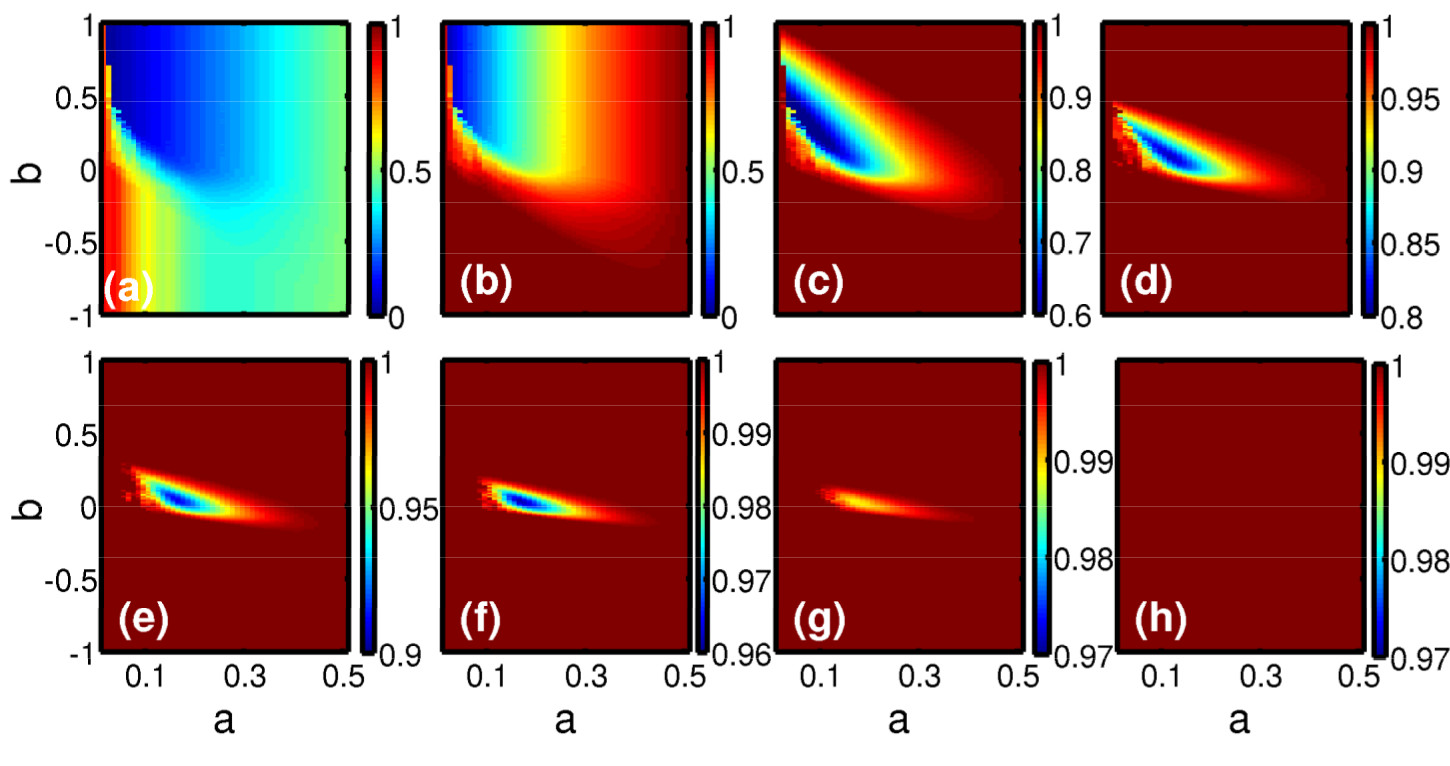}
\caption{\textbf{The eight accumulative indices $\eta_{1}$ (a), $\eta_{2}$ (b), $\eta_{3}$ (c), $\eta_{4}$ (d), $\eta_{5}$ (e), $\eta_{6}$ (f), $\eta_{7}$ (g), and $\eta_{8}$ (h) as a function of $a$ and $b$ for  Moore lattice.} The other parameters are token as $n_{s}=2$ and $L=101$. Each data point is obtained by averaging $100$ independent realizations. In the wide parameter regions for positive persistence, especially nearby the critical boundary (threshold), only $\eta_{i}>80\%$ ($i\geq \frac{\langle k \rangle}{2}$) rather than $\eta_{j}$ ($j<3$), where transmission events $E_{i}(t)$ ($i> 2$) contribute more than $80\%$ of the scale of message spreading (i.e., the spreading thus reaches a saturation level). It also implies that the majority of the population accept the message as truth only if the message is repeated more than three times in their ears. Therefore, the results with positive nearby the threshold also provide the evidence for the real phenomenon ``Three men make a tiger" (or ``A lie, if repeated often enough, will be accepted as truth").
}
\label{msatu2}
\includegraphics[width=\textwidth]{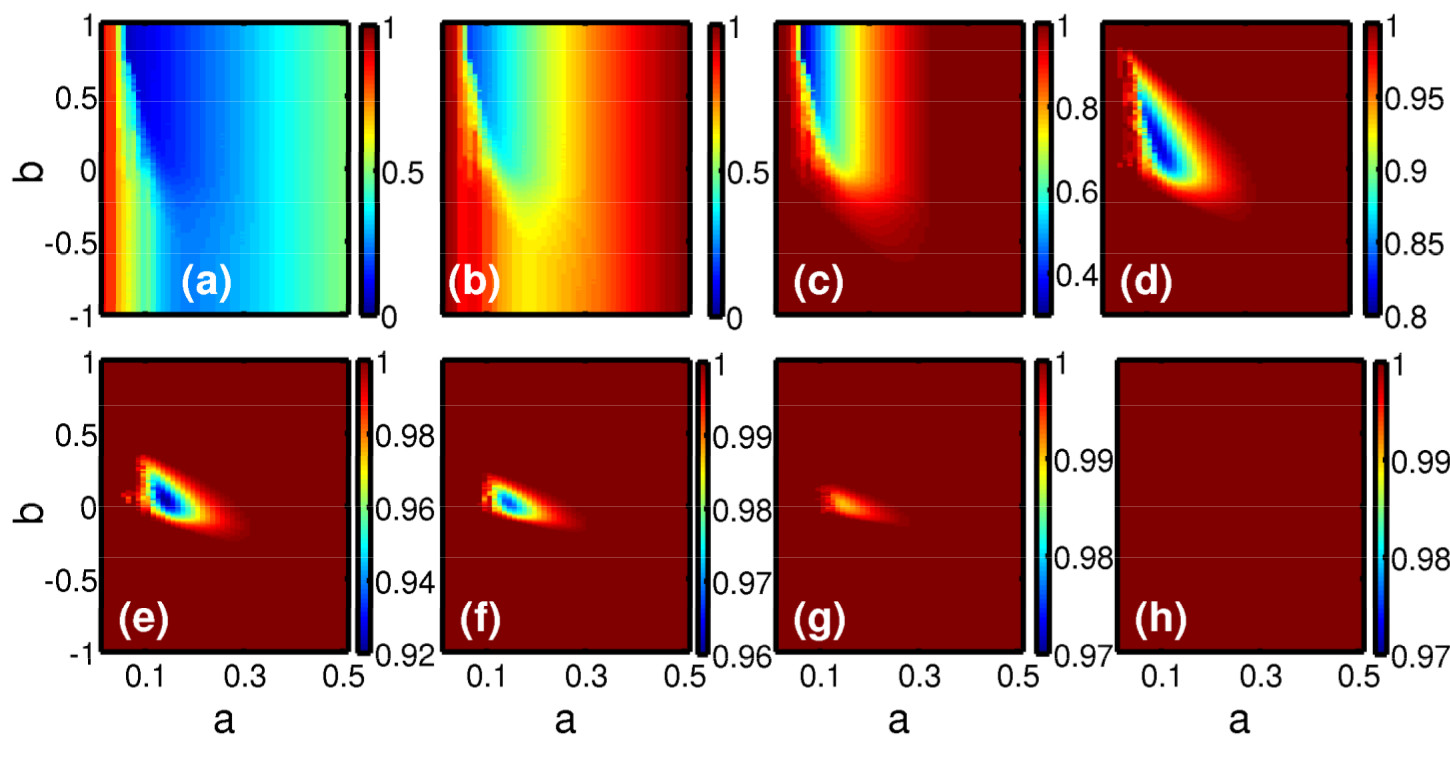}
\caption{\textbf{The eight accumulative indices $\eta_{1}$ (a), $\eta_{2}$ (b), $\eta_{3}$ (c), $\eta_{4}$ (d), $\eta_{5}$ (e), $\eta_{6}$ (f), $\eta_{7}$ (g), and $\eta_{8}$ (h) as a function of $a$ and $b$ for  Moore lattice.} The other parameters are token as $n_{s}=3$ and $L=101$. Each data point is obtained by averaging $100$ independent realizations. In the wide parameter regions for positive persistence, especially nearby the critical boundary (threshold), only $\eta_{i}>80\%$ ($i\geq \frac{\langle k \rangle}{2}$) rather than $\eta_{j}$ ($j<3$), where transmission events $E_{i}(t)$ ($i> 2$) contribute more than $70\%$ of the scale of message spreading (i.e., the spreading thus reaches a saturation level). It also implies that the majority of the population accept the message as truth only if the message is repeated more than three times in their ears. Therefore, the results with positive nearby the threshold also provide the evidences for the real phenomenon ``Three men make a tiger" (or ``A lie, if repeated often enough, will be accepted as truth").
}
\label{msatu3}
\end{figure}

\begin{figure}[ht]
\centering
\includegraphics[width=\textwidth]{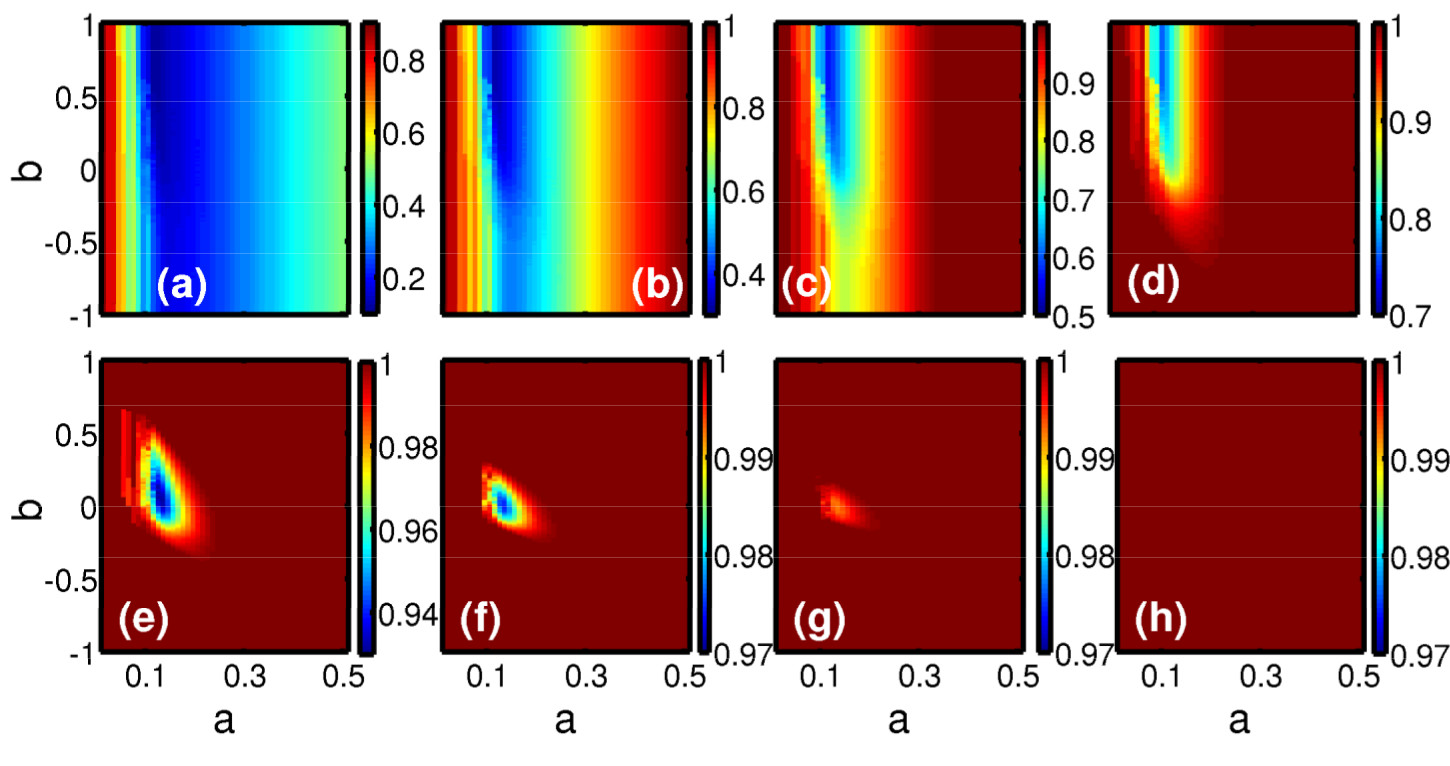}
\caption{\textbf{The eight accumulative indices $\eta_{1}$ (a), $\eta_{2}$ (b), $\eta_{3}$ (c), $\eta_{4}$ (d), $\eta_{5}$ (e), $\eta_{6}$ (f), $\eta_{7}$ (g), and $\eta_{8}$ (h) as a function of $a$ and $b$ for  Moore lattice.} The other parameters are token as $n_{s}=4$ and $L=101$. Each data point is obtained by averaging $100$ independent realizations. In the wide parameter regions for positive persistence, especially nearby the critical boundary (threshold), only $\eta_{i}>70\%$ ($i\geq \frac{\langle k \rangle}{2}$) rather than $\eta_{j}$ ($j<3$), where transmission events $E_{i}(t)$ ($i> 2$) contribute more than $70\%$ of the scale of message spreading (i.e., the spreading thus reaches a saturation level). It also implies that the majority of the population accept the message as truth only if the message is repeated more than three times in their ears. Therefore, the results with positive nearby the threshold also provide the evidences for the real phenomenon ``Three men make a tiger" (or ``A lie, if repeated often enough, will be accepted as truth").
}
\label{msatu4}
\includegraphics[width=\textwidth]{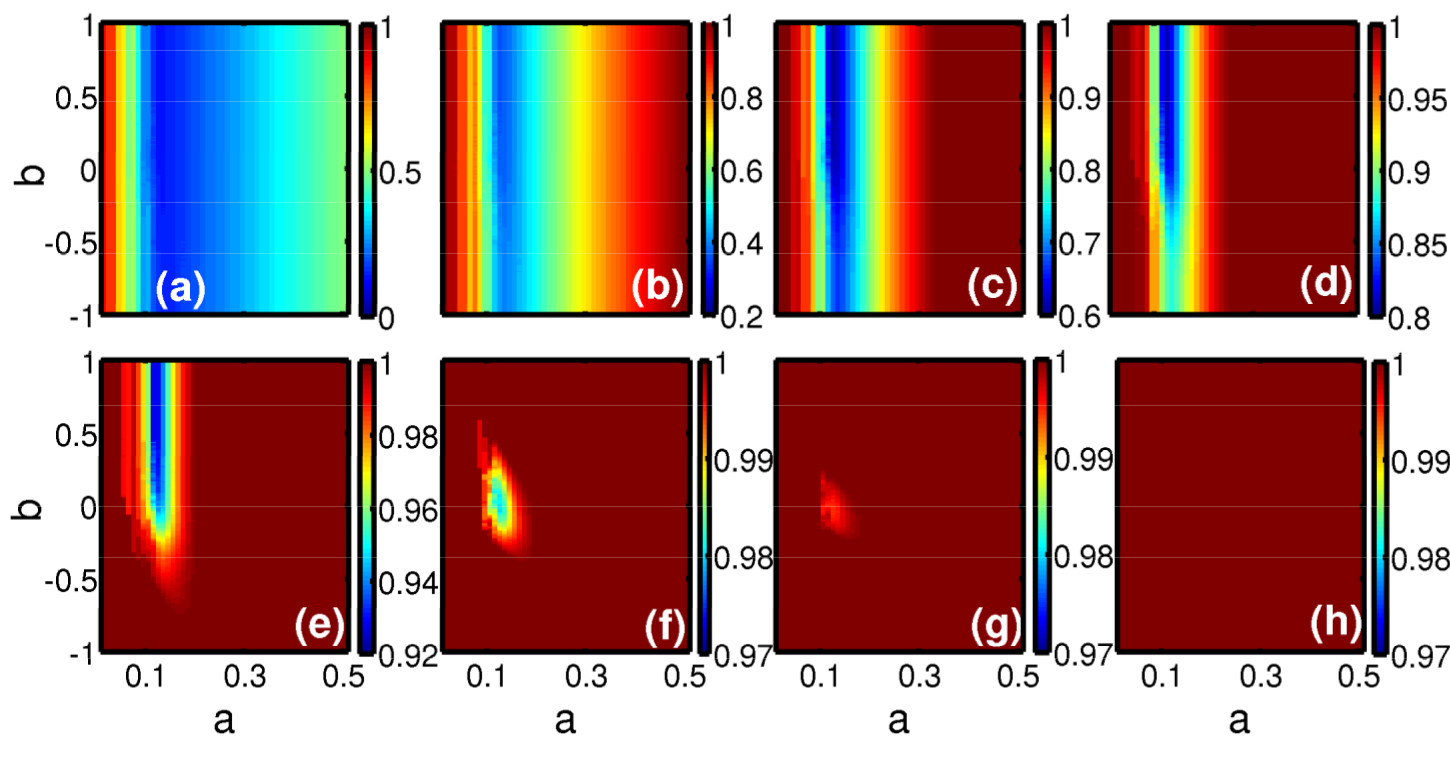}
\caption{\textbf{The eight accumulative indices $\eta_{1}$ (a), $\eta_{2}$ (b), $\eta_{3}$ (c), $\eta_{4}$ (d), $\eta_{5}$ (e), $\eta_{6}$ (f), $\eta_{7}$ (g), and $\eta_{8}$ (h) as a function of $a$ and $b$ for  Moore lattice.} The other parameters are token as $n_{s}=5$ and $L=101$. Each data point is obtained by averaging $100$ independent realizations. In the wide parameter regions for positive persistence, especially nearby the critical boundary (threshold), only $\eta_{i}>80\%$ ($i\geq \frac{\langle k \rangle}{2}$) rather than $\eta_{j}$ ($j<3$), where transmission events $E_{i}(t)$ ($i> 2$) contribute more than $70\%$ of the scale of message spreading (i.e., the spreading thus reaches a saturation level). It also implies that the majority of the population accept the message as truth only if the message is repeated more than three times in their ears. Therefore, the results nearby the threshold also provide the evidences for the real phenomenon ``Three men make a tiger" (or ``A lie, if repeated often enough, will be accepted as truth").
}
\label{msatu5}
\end{figure}

\begin{figure}[ht]
\centering
\includegraphics[width=\textwidth]{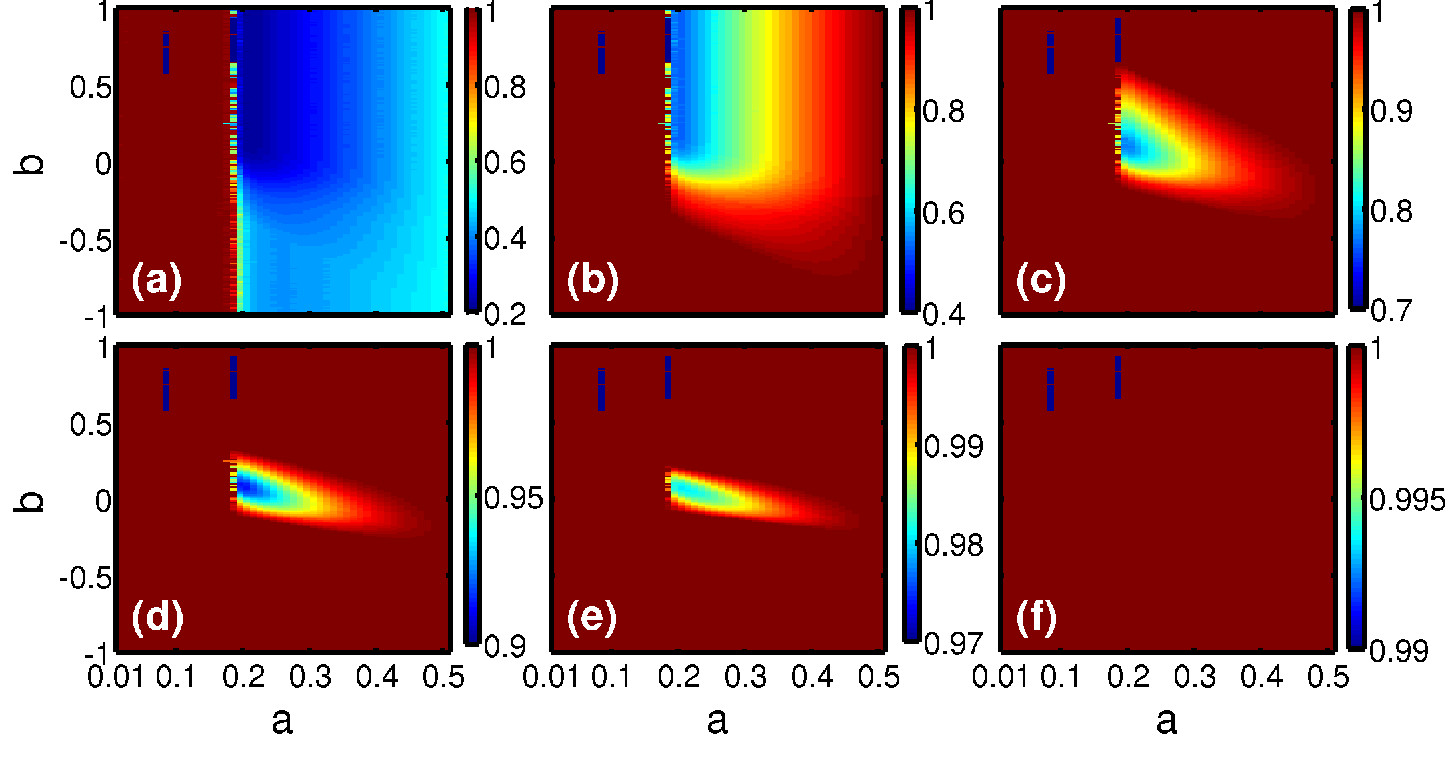}
\caption{\textbf{The six accumulative indices $\eta_{1}$ (a), $\eta_{2}$ (b), $\eta_{3}$ (c), $\eta_{4}$ (d), $\eta_{5}$ (e), and $\eta_{6}$ (f) as function of $a$ and $b$ for RRs with $\langle k \rangle=6$.} The other parameters are token as $n_{s}=5$ and $N=10000$. Each data point is obtained by averaging $100$ independent realizations. In the wide parameter regions for positive persistence, especially nearby the critical boundary (threshold), only $\eta_{i}>80\%$ ($i\geq \frac{\langle k \rangle}{2}$) rather than $\eta_{j}$ ($j<3$), , where transmission events $E_{i}(t)$ ($i> 2$) contribute more than $60\%$ of the scale of message spreading (i.e., the spreading thus reaches a saturation level). It also implies that the majority of the population accept the message as truth only if the message is repeated more than three times in their ears. Therefore, the results with positive persistence nearby the threshold also provide the evidences for the real phenomenon ``Three men make a tiger" (or ``A lie, if repeated often enough, will be accepted as truth").
}
\label{r6satu2}
\includegraphics[width=\textwidth]{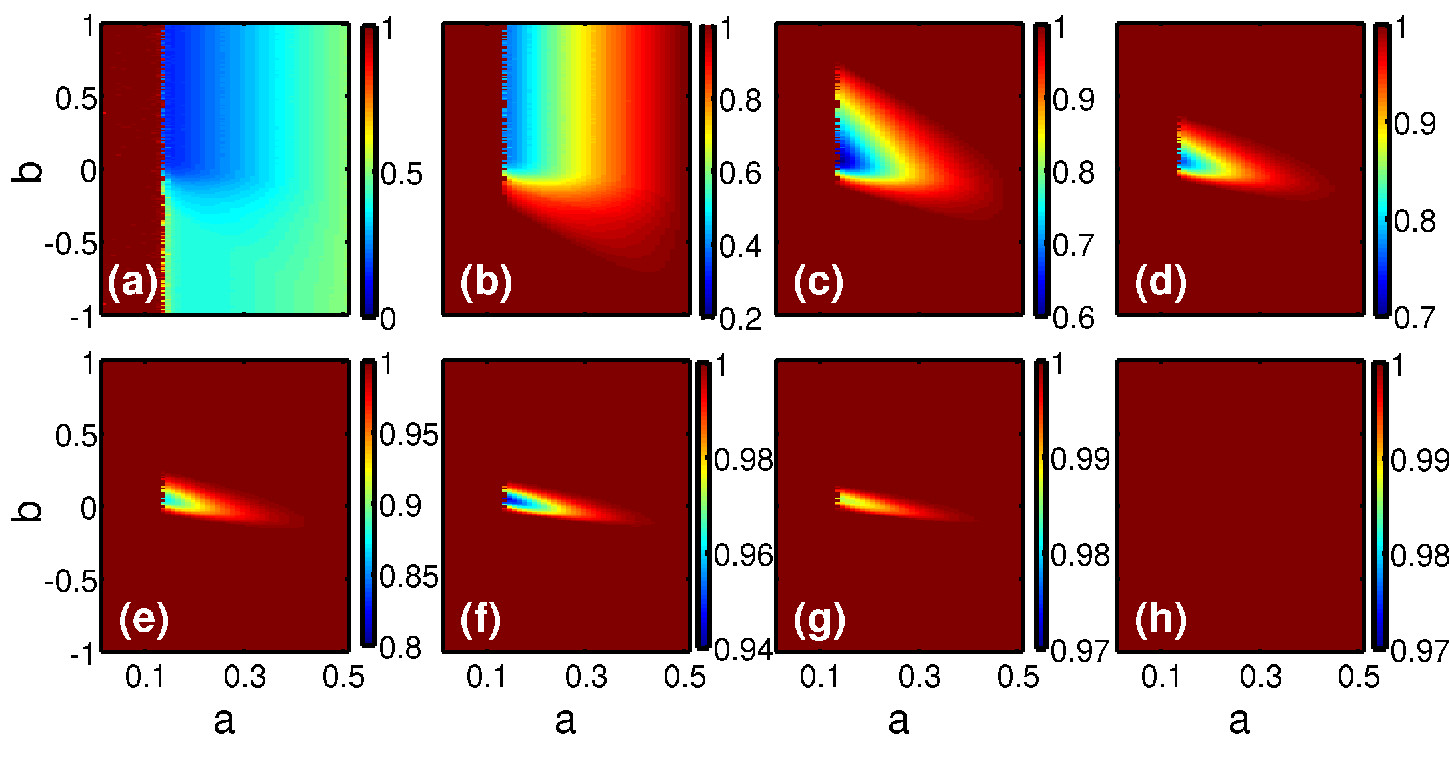}
\caption{\textbf{The eight accumulative indices $\eta_{1}$ (a), $\eta_{2}$ (b), $\eta_{3}$ (c), $\eta_{4}$ (d), $\eta_{5}$ (e), $\eta_{6}$ (f), $\eta_{7}$ (g), and $\eta_{8}$ (h) as function of $a$ and $b$ for RRs with $\langle k \rangle=8$.} The other parameters are token as $n_{s}=5$ and $N=10000$. In the wide parameter regions for positive persistence, especially nearby the critical boundary (threshold), only $\eta_{i}>80\%$ ($i\geq \frac{\langle k \rangle}{2}$) rather than $\eta_{j}$ ($j<3$), where transmission events $E_{i}(t)$ ($i> 2$) contribute more than $60\%$ of the scale of message spreading (i.e., the spreading thus reaches a saturation level).  It also implies that the majority of the population accept the message as truth only if the message is repeated more than three times in their ears. Therefore, the results with positive persistence nearby the threshold also provide the evidences for the real phenomenon ``Three men make a tiger" (or ``A lie, if repeated often enough, will be accepted as truth").
}
\label{r8satu2}
\end{figure}

\label{be}
\begin{figure}[ht]
\centering
\includegraphics[width=\textwidth]{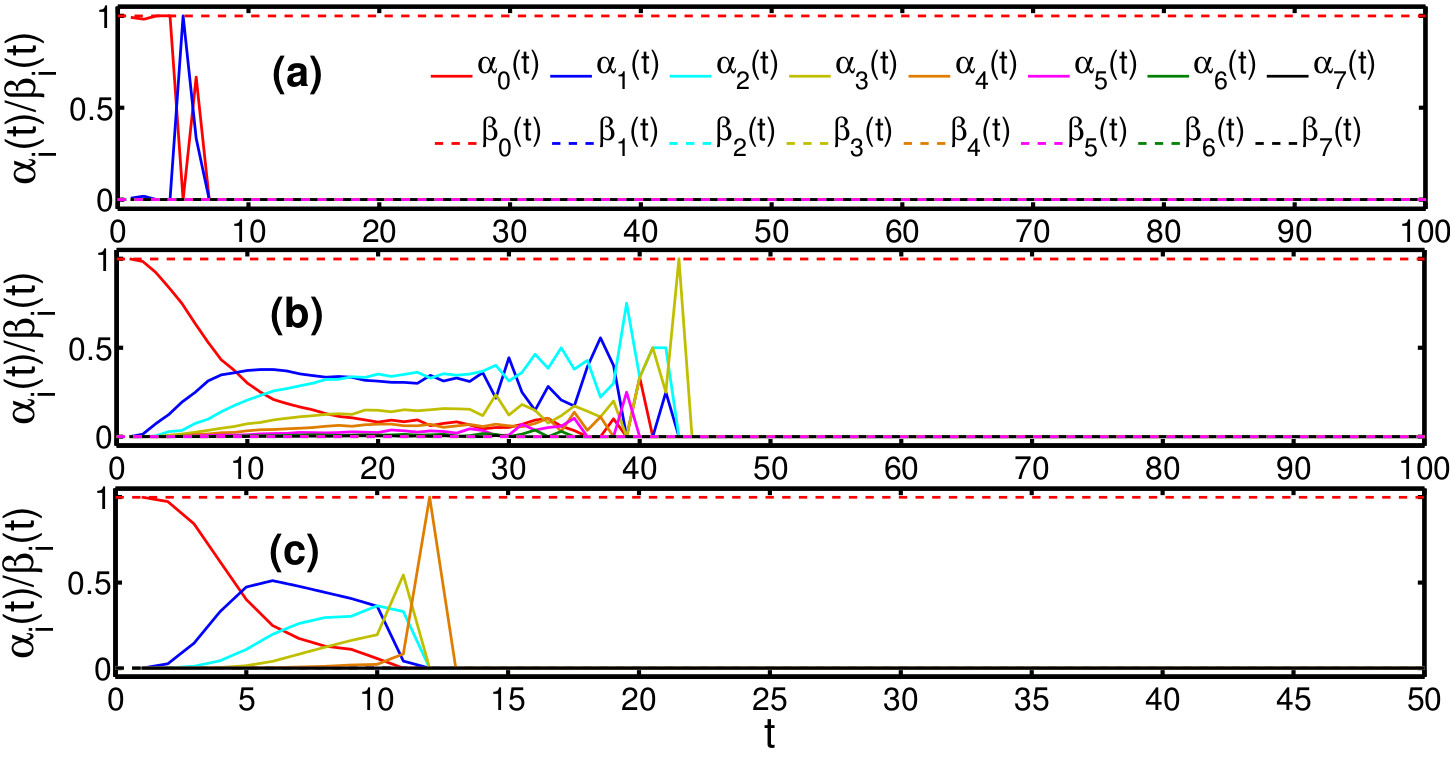}
\caption{\textbf{The evolution of proportions of the transmission events.} The evolution of indices $\alpha_{i}(a,b,t)$ from simulation (solid lines) and $\beta_{i}(a,b,t)$ from predications of percolation theory (dashed lines) are presented for SF with $\left\langle k\right\rangle=8$. Three difference cases are considered here: (a) the information vanishes for $a=0.10$, $b=0.20$, (b) it outbreaks for $a=0.19$, $b=0.20$ and prevails for $a=0.30$, $b=0.20$ (c). The other parameters are token as $n_{s}=5$ and $N=10000$. It can be observed that the occurrences of all the transmission events fail to last simultaneously and stably in the whole spreading process, departing from what exhibited in Fig.~S\ref{malpharound2} for moore lattice. The time correlations among the transmission events can thus be neglected in estimating the critical behaviors of the message spreading.
}
\label{baalpharound}
\end{figure}

\begin{figure}[ht]
\centering
\includegraphics[width=\textwidth]{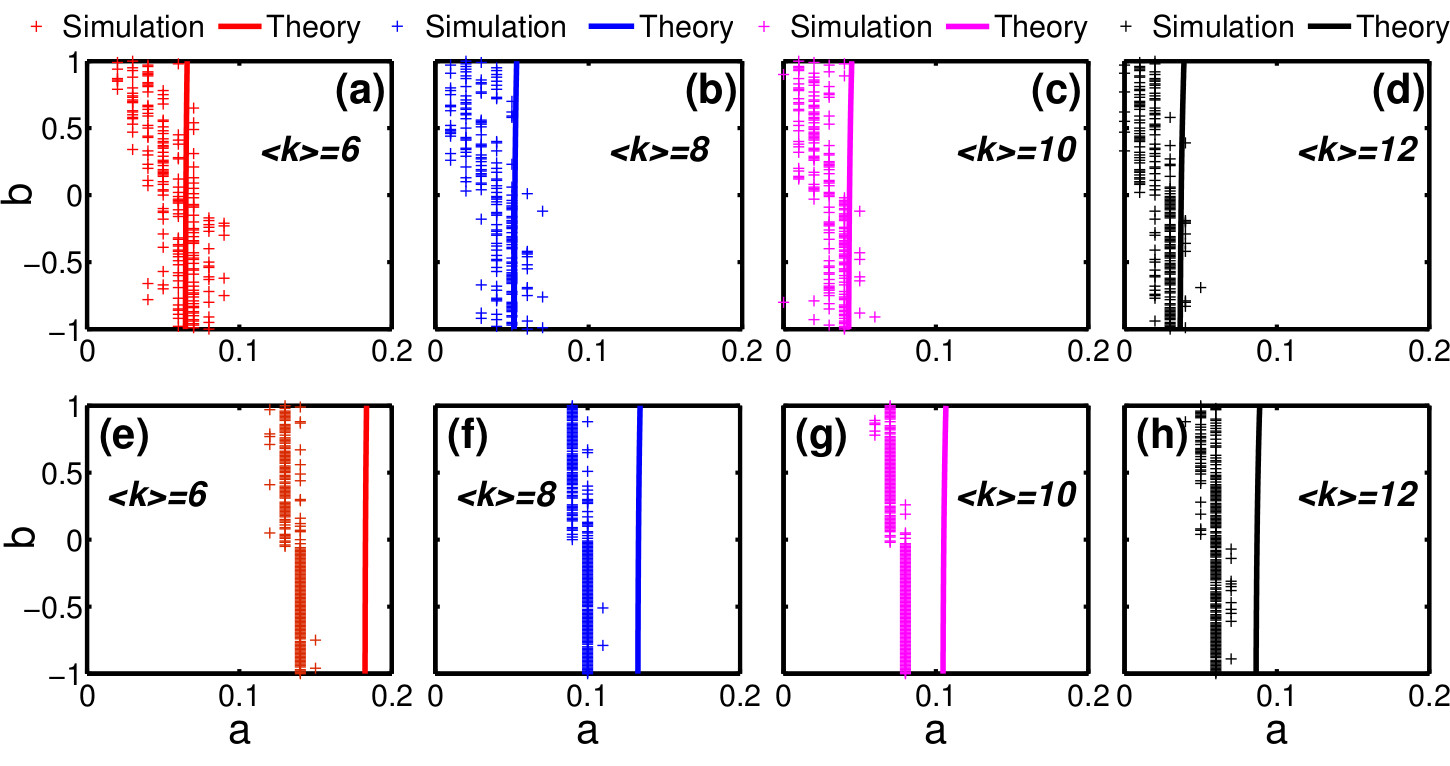}
\caption{\textbf{Locus of thresholds of message diffusion on SF networks and ER networks}. Respectively, analytical solutions (solid lines) for four SF networks (top panel) and four ER networks (bottom panel) with different average degrees are plotted to compare with the corresponding exact numerical data (markers). The other parameters are token as $n_{s}=2$ and $N=10000$. Each numerical data point is obtained by averaging $1000$ independent realizations. And (a)(e) $\left\langle k\right\rangle=6$, (b)(f) $\left\langle k\right\rangle=8$, (c)(h) $\left\langle k\right\rangle=10$ and (d)(i) $\left\langle k\right\rangle=12$. Both the simulations and the analytical predications show that the critical behaviors of the spreading are dominated by the stickiness of the message ($b$).}
\label{ae8thre}
\end{figure}

\begin{figure}[ht]
\centering
\includegraphics[width=\textwidth]{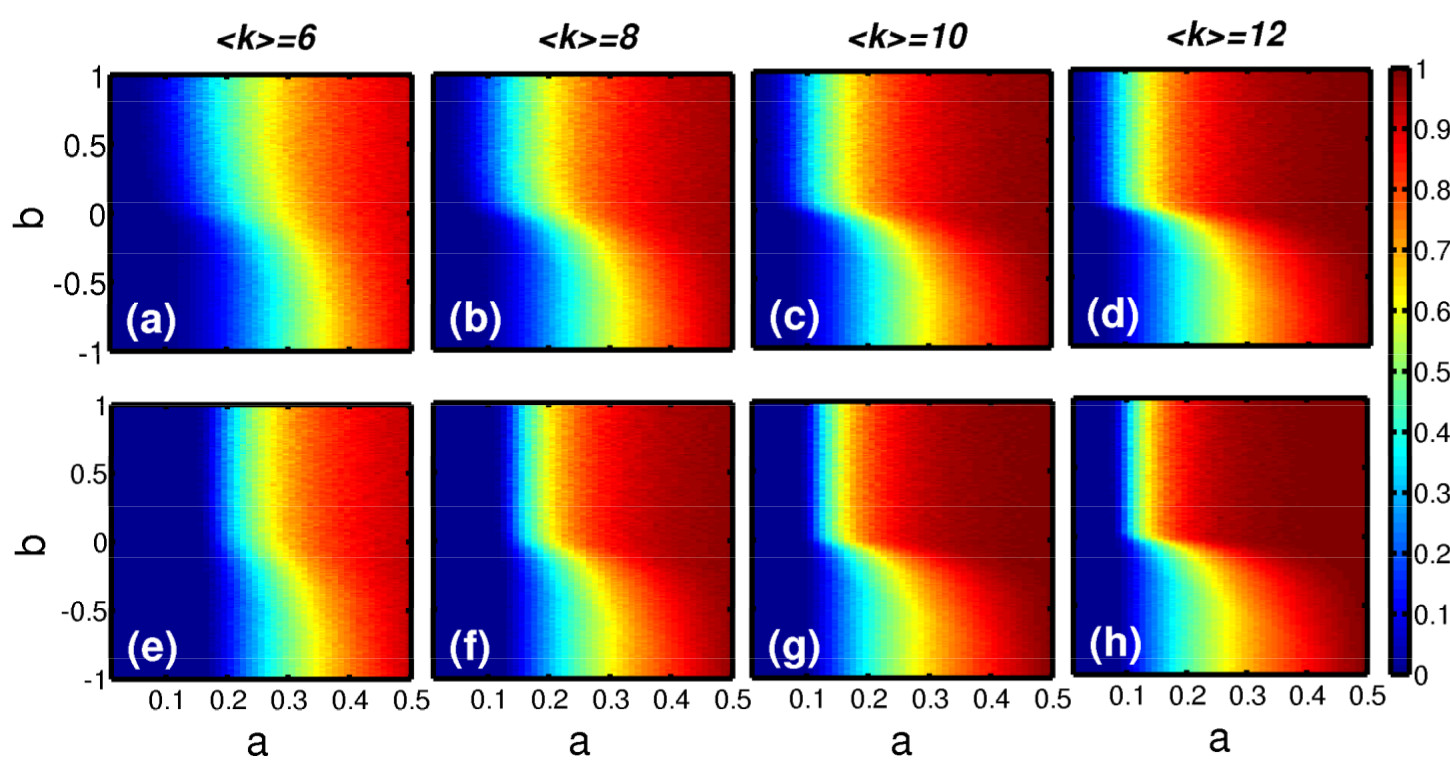}
\caption{\textbf{The densities of recovered individuals as function of $a$ and $b$,} on SF networks (top panel) and Erdos-Renyi networks (bottom panel) with different average degrees. The other parameters are token as $n_{s}=2$ and $N=10000$. Each numerical data point is obtained by averaging $100$ independent realizations. Precisely, (a)(e) $\left\langle k\right\rangle =6$, (b)(f) $\left\langle k\right\rangle =8$, (c)(h) $\left\langle k\right\rangle=10$ and (d)(i) $\left\langle k\right\rangle =12$. Although the critical behaviors of the message spreading on SF and ER are dominated by the stickiness, the persistence of the message (i.e., $b$) has a reasonable impact on the sizes of message spreading, especially at the parameter regions nearby $b=0$.
}
\label{beinfor}
\end{figure}

\begin{figure}[ht]
\centering
\includegraphics[width=\textwidth]{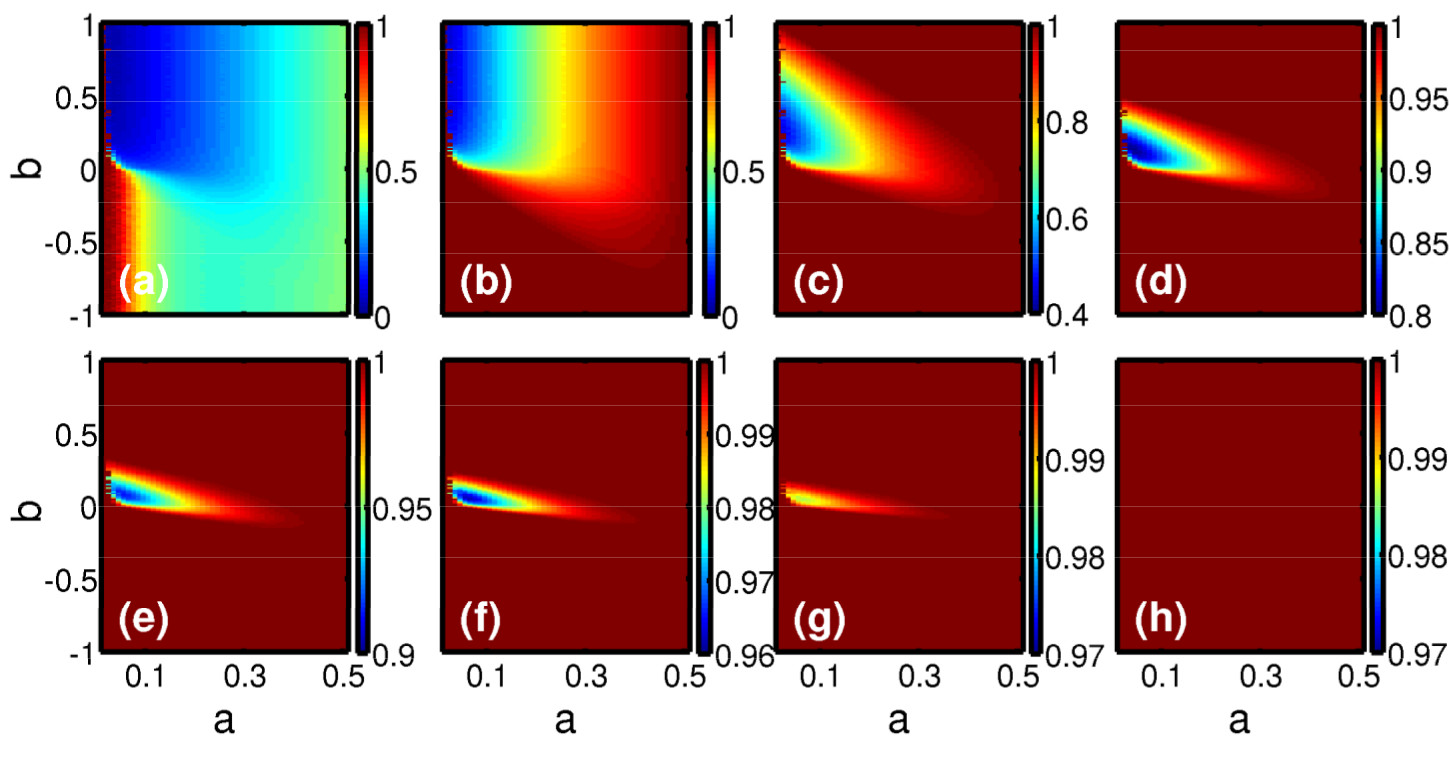}
\caption{\textbf{The eight accumulative indices $\eta_{1}$ (a), $\eta_{2}$ (b), $\eta_{3}$ (c), $\eta_{4}$ (d), $\eta_{5}$ (e), $\eta_{6}$ (f), $\eta_{7}$ (g) $\eta_{8}$ (h) as function of $a$ and $b$ for SF networks with $\langle k \rangle=8$.} The other parameters are token as $n_{s}=2$ and $N=10000$. Each numerical data point is obtained by averaging $100$ independent realizations. In the wide parameter regions for positive persistence, especially nearby the critical boundary (threshold), only $\eta_{i}>80\%$ ($i\geq \frac{\langle k \rangle}{2}$) rather than $\eta_{j}$ ($j<3$), where transmission events $E_{i}(t)$ ($i> 2$) contribute more than $80\%$ of the scale of message spreading (i.e., the spreading thus reaches a saturation level).  It also implies that the majority of the population accept the message as truth only if the message is repeated more than three times in their ears. Therefore, the results for $b>0$ nearby the threshold also provide the evidences for the real phenomenon ``Three men make a tiger" (or ``A lie, if repeated often enough, will be accepted as truth"). Similar results are obtained for SF networks with the same average degree.
}
\label{er6satu}
\includegraphics[width=\textwidth]{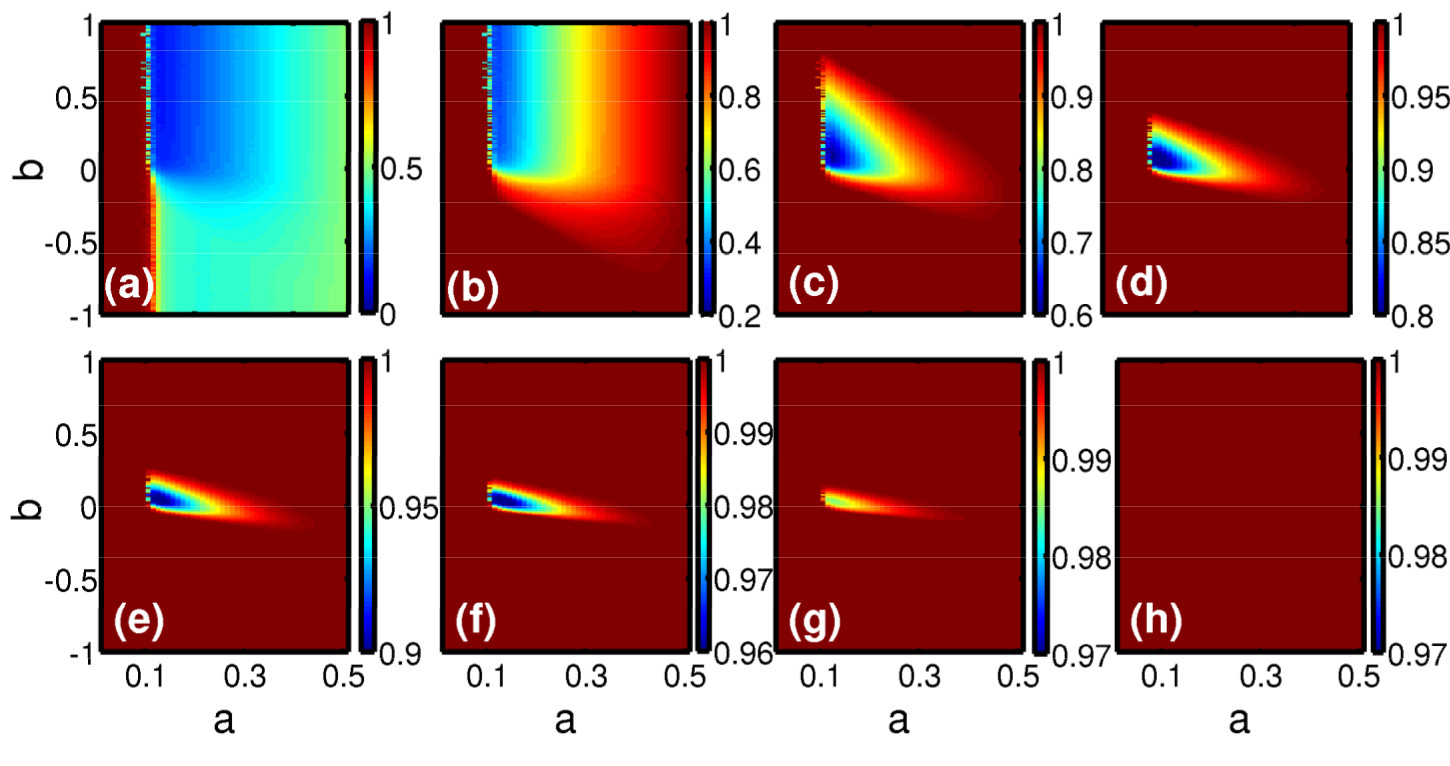}
\caption{\textbf{The eight accumulative indices $\eta_{1}$ (a), $\eta_{2}$ (b), $\eta_{3}$ (c), $\eta_{4}$ (d), $\eta_{5}$ (e), $\eta_{6}$ (f), $\eta_{7}$ (g) $\eta_{8}$ (h) as function of $a$ and $b$ for ER networks with $\langle k \rangle=6$.} The other parameters are token as $n_{s}=2$ and $N=10000$. Each numerical data point is obtained by averaging $100$ independent realizations. In the wide parameter regions for positive persistence, especially nearby the critical boundary (threshold), only $\eta_{i}>80\%$ ($i\geq \frac{\langle k \rangle}{2}$) rather than $\eta_{j}$ ($j<3$), where transmission events $E_{i}(t)$ ($i> 2$) contribute more than $70\%$ of the scale of message spreading (i.e., the spreading thus reaches a saturation level).  It also implies that the majority of the population accept the message as truth only if the message is repeated more than three times in their ears. Therefore, the results for $b>0$ nearby the threshold also provide the evidences for the real phenomenon ``Three men make a tiger" (or ``A lie, if repeated often enough, will be accepted as truth"). Similar results are obtained for SF networks with the same average degree.
}
\label{ba6satu}
\end{figure}

\end{document}